\documentclass[sigconf]{acmart}
\renewcommand\footnotetextcopyrightpermission[1]{} 
\newcommand\vldbdoi{XX.XX/XXX.XX}

\newcommand\vldbavailabilityurl{}

\usepackage{balance}
\usepackage{graphicx}
\usepackage{xspace}
\usepackage[section]{placeins}
\usepackage{url}
\usepackage{amsfonts}
\usepackage{amsmath}
\usepackage{paralist}
\usepackage[ruled]{algorithm2e}
\usepackage{adjustbox}
\usepackage{caption}
\usepackage{subcaption}
\usepackage{multirow}
\usepackage[page]{appendix}

\newcommand{\papername}{slo-admission-control}

\newcommand{\eg}{\emph{e.g.,}\xspace}
\newcommand{\ie}{\emph{i.e.,}\xspace}

\newcommand{\EQ}{{Eq.}}

\newcommand{\liquid}{{LIquid}\xspace}
\newcommand{\bouncer}{{Bouncer}\xspace}
\newcommand{\mqwt}{{MaxQWT}\xspace}
\newcommand{\mql}{{MaxQL}\xspace}
\newcommand{\af}{{AcceptFraction}\xspace}

\newcommand{\aallow}{{acceptance-allowance}\xspace}
\newcommand{\Aallow}{{Acceptance-allowance}\xspace}
\newcommand{\htu}{{helping-the-underserved}\xspace}
\newcommand{\Htu}{{Helping-the-underserved}\xspace}

\newcommand{\wrk}{\texttt{wrk2}\xspace}

\begin{document}

\title{
Admission Control with
Response Time Objectives 
for Low-latency Online Data Systems
}

\author{Hao Xu}
\affiliation{
  \institution{LinkedIn Corporation}
  \city{Mountain View}
  \state{CA}
  \country{USA}
}

\author{Juan A. Colmenares}
\affiliation{
  \institution{LinkedIn Corporation}
  \city{Mountain View}
  \state{CA}
  \country{USA}
}


\begin{abstract}
  To provide quick responses to users, Internet companies rely on online data systems able to answer
  queries in milliseconds. 
  These systems employ complementary overload management techniques to ensure they provide a
  continued, acceptable service throughout traffic surges, where ``acceptable'' partly means that
  serviced queries meet or track closely their response time objectives.
  Thus, in this paper we present \emph{\bouncer}, 
  an admission control policy aimed to keep admitted queries under or near their
  service level objectives (SLOs) on percentile response times.
  It computes inexpensive estimates of percentile response times for every incoming query and
  compares the estimates against the objective values to decide whether to accept or reject the
  query.
  \bouncer allows assigning separate SLOs to different classes of queries in the workload,
  implements early rejections to let clients react promptly and to help data systems avoid doing
  useless work, 
  and complements other load shedding policies that guard systems from exceeding their capacity.
  Moreover, we propose two \emph{starvation avoidance strategies} that supplement \bouncer's basic
  formulation and prevent query types from receiving no service (starving).

  We evaluate \bouncer and its starvation-avoiding variants against other policies in simulation and
  on a production-grade in-memory distributed graph database.
  Our results show that \bouncer and its variants allow admitted queries to meet or stay close to
  the SLOs when the other policies do not.
  They also report fewer overall rejections, a small overhead, and with the given latency SLOs,
  they let the system reach high utilization. 
  In addition, we observe that the proposed strategies can stop query starvation, but at the expense of a
  modest increase in overall rejections and causing SLO violations for serviced requests;
  however, our results suggest that the violation counts may be acceptable in practice.
\end{abstract}


\maketitle



\section{Introduction}
\label{sec:intro}

Social media, e-commerce, and other Internet companies rely on low-latency online data
systems~\cite{dynamodb2007,tao2013,spanner2013,analyticdb2019,cosmosdb,liquidteam}  
to provide quick responses to their users.
These large-scale systems answer queries from multiple clients demanding millisecond-scale response
times (\eg 1ms--100ms), and each of their servers can receive tens or even a few hundreds of
thousand queries per second (QPS).
The queries are normally of various types with different complexity and latency characteristics.
Customers usually want to know the response times a system can offer to their queries to ensure
their latency requirements are met.
To set clear performance goals for themselves and realistic expectations for customers, system
operators establish service level objectives (SLOs)~\cite{jones2017} on query response times. 
These latency SLOs are typically defined in terms of percentiles (\eg p50=10ms and p90=60ms), and
having separate SLOs for different classes of queries is common.

With users driving the inbound traffic, they can cause surges in which numerous simultaneous
requests exceed the maximum load a serving system can handle with the provisioned
resources~\cite{lefebvre2001,kim2018,purnell2020}.
Besides legitimate, well-intended users, malicious agents launching distributed denial of service
(DDoS) attacks~\cite{jonker2017,kopp2021,toh2022} can exacerbate the situation. 
Even with the inbound traffic below the tolerable limit, there are other reasons for online data
systems to become overloaded. 
Unplanned reduction in the system's capacity\footnote{
\ie the maximal traffic load, in QPS, that the system is able to sustain.
}
can result from network outages, node failures, configuration changes, and software bugs.
Moreover, Internet companies typically conduct live traffic load
tests~\cite{kraken2016,trafficshift2017} on a regular basis (\eg daily) to confirm that
their system infrastructure can handle the extra load when outages in one or more data centers
occur.

To ensure continued operation throughout possible load surges, online data systems 
(and their surrounding infrastructure) 
employ complementary overload management techniques~\cite{cuervo2017b,taxonomy2018} that tackle the
problem from different angles, such as  
load balancing~\cite{lewandowski2017,cuervo2017a,taiji2019}, 
per-client quotas and client-side throttling~\cite{cuervo2017b,doorman2016}, 
resource scheduling~\cite{colajanni1997,schroeder2006}, 
and 
graceful degradation~\cite{liu1994,mittal2016,blinkdb2013,ulrich2017}.
Another angle is \emph{admission control} 
(load
shedding)~\cite{chen2002session,welsh03,elnikety2004,quorum2005,qcop2010,activesla2011,zhang2014,ulrich2017,zhou2018overload,mittos2017,breakwater2020,sap_hana_rm2019,darling2022},
which is the focus of this paper.

Based on the system's load status, an admission control module decides to accept or reject client
requests so that the system continues providing an acceptable service when receiving excessive
traffic. 
Some admission control techniques (\eg~\cite{heiss1991,weikum1992,elnikety2004}) 
concentrate on allowing the system to serve a sustained throughput without exceeding its capacity,
protecting it from performance collapse. 
But, for online data systems ``acceptable service'' also means that admitted queries should meet or
at least track closely the latency SLOs.
Hence, in this paper we present \emph{\bouncer} (\S\ref{sec:slo_policy}), 
a measurement-based admission control policy whose main goal is to \emph{keep serviced queries under
or close to their latency SLOs} defined in terms of percentile response times.
It implements \emph{early rejections}~\cite{ulrich2017,mittos2017,breakwater2020} to 
help data systems avoid doing useless work and give clients more flexibility to failover promptly. 
It is also designed to \emph{complement} other admission control policies that keep systems from
exceeding their capacity.
For workloads with diverse queries, \bouncer allows assigning SLOs per query type.

The \emph{main challenge} here lies in estimating percentile response times quickly and effectively
for query-by-query acceptance decision making.
\bouncer computes estimates of percentile response times for each incoming query by combining
\emph{inexpensive approximations} of queue wait time and processing time from recent query executions, and
compares the estimates with the objective values to decide to accept or reject the query. 
An additional issue is that \bouncer's basic formulation may systematically deny service to query
types, especially under heavy load.
To prevent query types from receiving no service (starving), we supplement \bouncer with two
alternative \emph{starvation avoidance strategies} (\S\ref{sec:query_starvation}).

We evaluate \bouncer and its starvation-avoiding variants against several in-house policies, in
simulation (\S\ref{sec:simulation_results}) and on \liquid, a production-grade in-memory distributed
graph database (\S\ref{sec:impeller_results}).
We describe the in-house policies in \S\ref{sec:other_policies} and provide an overview of
\liquid~\cite{liquid20a,liquid20b} in \S\ref{sec:liquid}.
We show that \bouncer and its variants are able to make effective decisions, and allow admitted
queries to meet or stay near the latency objectives when the other policies do not.
They report 
fewer overall rejections (15\%$\sim$30\%) than the other policies,
a small overhead (mean=18$\mu$s) for millisecond-scale queries, 
and with the given latency SLOs, they let the system reach high utilization.
Further, we show that the proposed strategies can stop query starvation, but at the expense of a
modest increase in overall rejections and causing SLO violations for serviced requests; however, our
results on \liquid graph database suggest the violation counts may be acceptable in practice. 

To sum up, the contributions of this paper are:
\begin{compactitem}
\item
  an \emph{admission control policy}, called \bouncer, that  
  promptly rejects queries expected to surpass their response time objectives and 
  lets admitted queries meet or stay near theirs; 
\item
  two \emph{starvation avoidance strategies} --
  \emph{\aallow} (\S\ref{sec:acceptance_allowance}) and
  \emph{\htu} (\S\ref{sec:help_underserved}) -- 
  that complement \bouncer and protect query types from systemic service denial;
  and
\item
  an \emph{extensive evaluation} of \bouncer and its starvation-avoiding variants in simulation and
  on \liquid (\ie a real system).
\end{compactitem}

The paper ends with the related work (\S\ref{sec:related_work}), our conclusions and future work
(\S\ref{sec:conc}). 
In addition, it provides a supplementary discussion about  
our plan to handle cold starts and traffic lulls in \bouncer (Appendix~\ref{sec:cold_start}) 
and 
practical considerations (Appendix~\ref{sec:considerations}) related to 
the selection of percentiles for the latency SLOs and
\bouncer's configuration effort.

\section{Motivation}
\label{sec:motivation}

In this section, we discuss the motivation behind the main design decisions in \bouncer.

\subsubsection*{\textbf{Early rejections}}
\label{sec:early_rejects}

Online data systems often operate as shared services at the bottom of a multi-tier architecture,
with microservices~\cite{newman2021} in the middle and API gateways~\cite{richardson2019}
at the top as entry points for external mobile or web client applications.
Top- and mid-tier services set expiration times (deadlines) for their requests to downstream
services, and give up on requests that time out to offer users timely, yet lower-quality responses. 
For example, suppose a mid-tier service sends simultaneous requests $r_{1}$, $r_{2}$ and $r_{3}$
with a 50ms deadline to give enough time to upstream services to produce their responses and the
mobile app to render the results. 
If $r_{3}$ takes longer than the deadline (let's say 55ms), then the service, after waiting until
the deadline expires, would ignore $r_{3}$'s response and may choose to only return the results from
$r_{1}$ and $r_{2}$ to its client.

When an online data system is under heavy load and operates nearly at peak capacity, its response
times increase and if no remediation is in place, queries may approach or reach their expiration
times, which can have several consequences.
\emph{First}, 
longer response times exacerbate the resource usage (\eg memory and threads) of mid-tier services
because the longer a query takes, the longer the upstream services in the call chain wait and hold
on to resources (until deadlines expire).
\emph{Second}, 
depending on how tight the deadlines are compared to the expected upper-bound response times,
upstream services may have little time to react to expired requests and produce alternative results.
Continuing with our example, if the mid-tier service is expected to respond in no more than 60ms,
after waiting 50ms for $r_{3}$ it would only have 10ms to provide a substitute, degraded response. 
\emph{Third}, 
deadline expirations throughout the service call chains become more likely when queries take longer;
when queries time out at the data system or its responses expire on their way up to the users, 
the system unnecessarily spends memory, CPU, and other resources -- precious during overload -- and
any work it does to answer the queries goes to waste.

To alleviate the above issues, we adopt a \emph{fail-early-and-cheaply} approach~\cite{ulrich2017} to
admission control on the online data system.
\bouncer rejects queries expected to miss their latency objectives straight away after arrival (\ie
these queries never make it into the data system's queue).
By giving early rejections~\cite{mittos2017,breakwater2020}, it 
\begin{inparaenum}[1)]
\item
helps reduce resource usage of upstream services, 
\item
offers these services more flexibility to decide the next action to obtain alternative results, 
and
\item
lowers the chances of the data system doing useless work.
\end{inparaenum}
Further, on the ``cheaply'' side, \bouncer uses inexpensive 
response time estimates 
to make acceptance decisions. 

\subsubsection*{\textbf{Complementing capacity-centric policies}}
\label{sec:complement}

We design \bouncer to complement, rather than replace, 
admission control policies aimed to
guard data systems from exceeding their capacity. 
The reason is that we cannot easily guarantee that a system is protected against overloading by just
enforcing percentile response times because the level of protection depends on the chosen percentile
values. 
For example, suppose a system 
can answer individual queries of type \texttt{A} in 10ms on average, 
is currently dedicated to serve such queries, and  
can process 64 queries in parallel; 
the system can thus sustain 6,400 ($=1/0.01 \times 64$) of type-\texttt{A} queries per second. 
If we adopt the customer's latency requirement of $SLO_{p50}=100ms$ for the \texttt{A} queries,\footnote{
Rather than an $SLO_{p50}$ closer to 10ms, in anticipation of other query types being soon in production.}
then \bouncer will only reject queries \emph{after} the system has taken work beyond its capacity. 
Moreover, finding the right latency SLOs to ensure the system is protected becomes more challenging
when the workload includes queries of various types with different performance characteristics. 
By contrast, some capacity-centric policies (\eg~\cite{cherkasova1998} and \af in
\S\ref{sec:acceptrate}) just require setting a few parameters, such as the maximum
utilization threshold.

Note that being complementary to capacity-centric policies also means that \bouncer should
not prevent the system from processing queries at its full capacity. 
But, as illustrated in our example, that is not difficult to achieve with loosen latency
requirements.

\subsubsection*{\textbf{Query type awareness}}
\label{sec:querytype_awereness}

Online data systems normally serve workloads carrying queries of various types with different
complexity and latency characteristics. 
In graph databases, for example, simple edge queries, which return the vertices directly
connected to a given vertex, are usually fast while graph distance queries, which determine the
shortest distance between two vertices, can take longer.
We choose to differentiate query types in \bouncer 
as prior research shows that 
\begin{inparaenum}[1)]
\item
  a query's type often has a big impact on its response time~\cite{elnikety2004},
  and
\item
  admission control techniques oblivious to query types tend to reject more queries than needed
  because they cannot determine which queries are more beneficial to reject~\cite{qcop2010}.  
\end{inparaenum}

\section{The ``\bouncer'' Policy}
\label{sec:slo_policy}

\bouncer is an admission control policy that makes acceptance decisions based on the latency SLOs of
queries.
It estimates percentile response times for each incoming query and compares them with the target
percentile response times in the SLO to decide whether to accept or reject the query.

\begin{figure}[!tp]
\centering
\includegraphics[width=0.95\columnwidth]{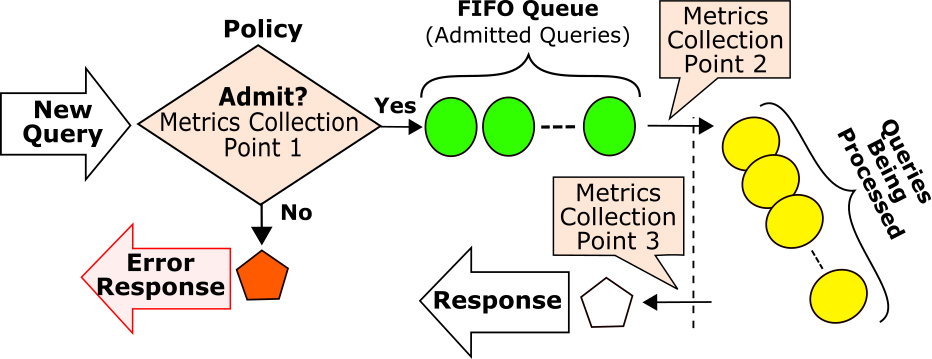}
\vspace{-.45cm}
\caption{Query processing with admission control and metrics collection.}
\label{fig:framework}
\end{figure}

\bouncer is built atop a software framework that resembles a stage in the staged event-driven
architecture (SEDA) with the multi-class overload controller~\cite{welsh03},
and it facilitates implementing thread-safe, highly-concurrent
policies.
Figure~\ref{fig:framework} depicts the query processing in the framework.
When a new query arrives, 
the policy examines it and, based on metrics gathered from recent executions, decides
to admit or reject it. 
If admitted, the query is inserted into the FIFO queue to wait for its turn to be processed;
otherwise, the policy drops it and instructs the server host to reply with an error response. 
A fixed number of query engine processes (or threads) dequeue the admitted queries and process each
independently.
The framework offers methods for the policies to record time intervals and metrics they need  
(\eg queue wait time, processing time, queue length, and rejection counts) 
at several points in the process:   
\emph{Point~1}, after the admission or rejection decision is made;  
\emph{Point~2}, after a query is dequeued for processing; 
and 
\emph{Point~3}, after a query has been processed and before the response is sent back to the client.

We assume that every request includes a short string indicating the type of the query it carries
(\eg part of the REST URL endpoint's path or the name of a datalog-like
rule~\cite{fodbs1995,liquid20b}). 
Thus, the policy is configured with strings denoting the query types and for each type, a latency
SLO with the target percentile response times; for example:
\texttt{"Fast"}:\{p50=10ms, p90=90ms\}, 
\texttt{"Slow"}:\{p50=60ms, p90=270ms\}, 
and  
\texttt{"default"}:\{p50=30ms, p90=400ms\}.\footnote{
These query types are merely for illustration purpose; in practice, they often have more sensible
names (\eg \texttt{GetFriends} and \texttt{GraphDistance}).
}
Note that \texttt{default} is a ``catch-all'' query type.

The time a query $Q$ spends in the FIFO queue is its \emph{queue wait time}; 
\ie $wt(Q) = t_{dequeued}(Q)-t_{enqueued}(Q)$.
We define a query $Q$'s \emph{processing time}, denoted as $pt(Q)$, as the time interval from the
instant at which $Q$ is pulled from the head of the queue to the instant at which $Q$ has been
fully processed and the response is ready to be sent back to the client; \ie   
$pt(Q) = t_{completed}(Q)-t_{dequeued}(Q)$.
Then, the \emph{response time} of a query $Q$ is:
\begin{equation}
  rt(Q) = wt(Q) + pt(Q) + \xi
  \label{eqn:response_time}  
\end{equation}
where $\xi$ is some additional time the server host takes 
(\eg in the network stack and operating system) 
to handle the query.
In our experience $\xi$ is often negligible and we assume $\xi=0$.

Separate query types often have different processing time distributions that vary over time. 
\bouncer adopts the natural approach of maintaining approximations for these distributions in
histograms, one per query type (including \texttt{default}); it periodically updates the histograms
at run time using a dual-buffer technique.\footnote{
  While one histogram is only read, a second histogram is being populated.
  At the end of a time interval the new and old histograms are swapped atomically, and the old
  histogram is reset before being populated again.
}

\begin{figure}[!tp]
\centering
\includegraphics[width=0.95\columnwidth]{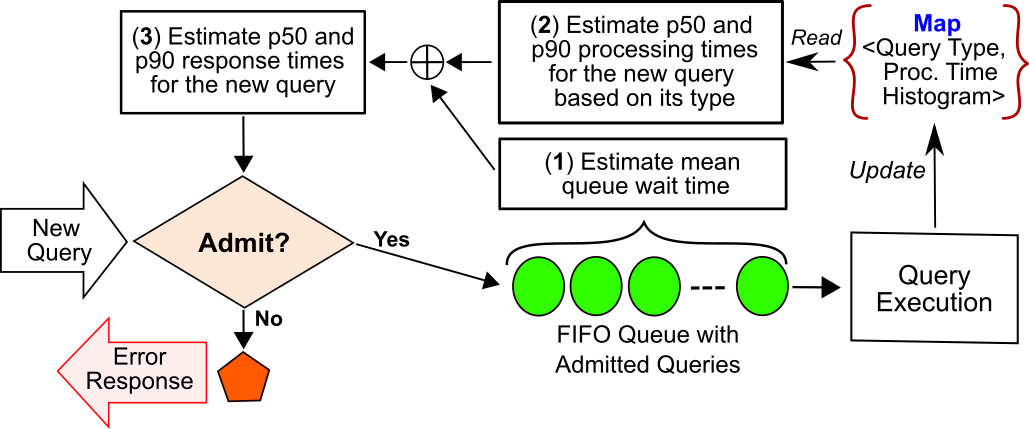}
\vspace{-.45cm}
\caption{Operation of \bouncer admission control policy.}
\label{fig:slo_diagram}
\end{figure}

Figure~\ref{fig:slo_diagram} illustrates \bouncer's operation.
For every query $Q$ that arrives at the host, it computes an estimate of the mean queue wait time
that $Q$ will experience as:
\begin{equation}
  ewt_{mean} = \frac{\sum_{type \in QT}{(count(type) \times pt_{mean}(type))}}{P}
  \label{eqn:ewt}
\end{equation}
where
$QT$ is the set of query types the policy recognizes, including \texttt{default};
$count(type)$ is the number of queries of the given $type$ currently in the queue;
$pt_{mean}(type)$ is the mean processing time for queries of the given $type$, obtained from the
corresponding histogram; and
$P$ is the number of query engine processes running on the host (\ie the level of task parallelism
for query processing).
\bouncer maintains per-type atomic counts of the queries currently in the queue, and updates the
counts as queries are enqueued and dequeued.
The histograms and query counts are stored in hash maps with query types as the keys for easy and
quick access.

\bouncer estimates the percentile response times of $Q$ as:
\begin{align}
  ert_{p50}(Q) &= ewt_{mean} + pt_{p50}(Type(Q)) \label{eqn:ert_p50} \\
  ert_{p90}(Q) &= ewt_{mean} + pt_{p90}(Type(Q)) \label{eqn:ert_p90}
\end{align}
where 
$Type(Q)$ returns the type of the query $Q$, and  
$pt_{p50}(type)$ and  $pt_{p90}(type)$ are the 50th- and 90th-percentile processing times for
queries of the given $type$, obtained from the corresponding histograms.
Finally, after calculating $ert_{p50}$ and $ert_{p90}$, 
\bouncer decides to accept or reject $Q$ according to Algorithm~\ref{alg:slo_decision}.

\begin{algorithm}[!tp]
 	\small
  \lIf{($ert_{p50}(Q) > SLO_{p50}(Q)$ \texttt{||}  $ert_{p90}(Q) > SLO_{p90}(Q)$)}{Reject}
  \lElse{Accept}
  \caption{Decision on incoming query $Q$.}
  \label{alg:slo_decision}
\end{algorithm}

Note that Equations~\ref{eqn:ewt}, \ref{eqn:ert_p50}, and \ref{eqn:ert_p90}, rather than precise
expressions, are inexpensive estimations we adopt to keep the policy's overhead low. 
This trade-off of accuracy for speed is of practical importance because the computations \bouncer
does are in the critical path of the queries.   
Our experiments (\S\ref{sec:eval}) show, however, that the policy is effective despite this accuracy
loss.
Moreover, this policy formulation can be easily modified to 
support SLOs with other percentile response times (\eg p99) in lieu of or in addition to p50 and p90,  
and 
to adopt different logical expressions for acceptance decision making in Algorithm~\ref{alg:slo_decision}. 
The evaluation of alternative formulations for \bouncer is left as future work, but
Appendix~\ref{sec:considerations} offers some practical considerations about selecting percentiles
for latency SLOs.

Also note that the \bouncer policy, as formulated above, suffers from two problems. 
The first one is \emph{query starvation} since the policy may systematically deny service to query types.
The second problem is that \bouncer is susceptible to \emph{cold starts} and \emph{lulls in traffic} as its
histograms start in a blank state and may become empty for query types with intermittent patterns. 
We address query starvation in the next section, but leave the issue of cold starts and traffic
lulls as future work and just discuss our plan in Appendix~\ref{sec:cold_start}.

\section{Avoiding Query Starvation}
\label{sec:query_starvation}

\begin{figure}[!tbp]
  \centering
  \includegraphics[width=0.75\columnwidth]{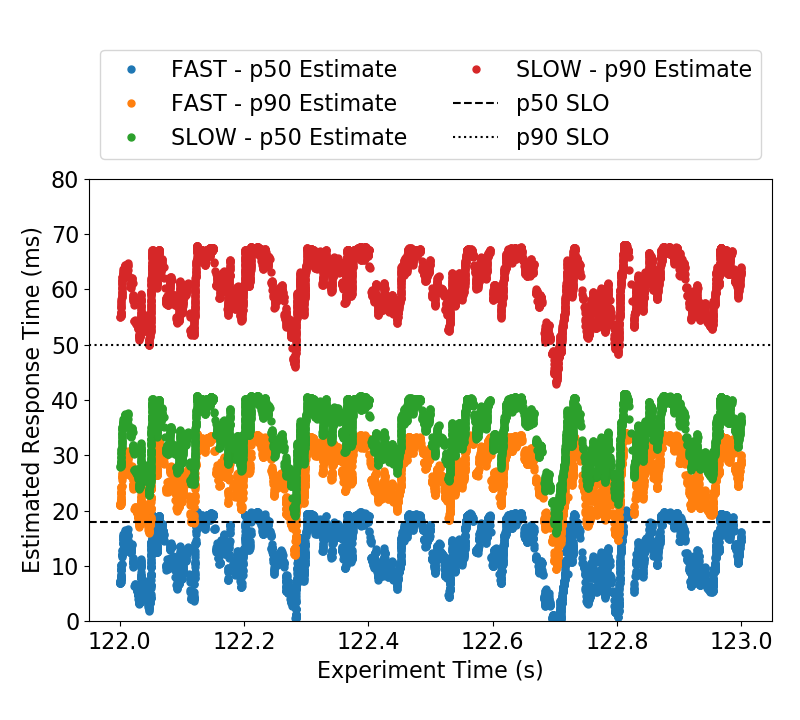}
  \vspace{-.60cm}
  \caption{
    Example of query starvation. 
    \small{
		The latency SLO is the same for \texttt{FAST} and \texttt{SLOW} queries: $SLO_{p50}=18ms$
    and $SLO_{p90}=50ms$ (dotted lines).
    Around $99\%$ of \texttt{SLOW} queries are rejected, while less than $10\%$ of \texttt{FAST}
    queries are rejected.
		}
  }
  \label{fig:query_starvation_example}
\end{figure}

Query types frequently have latency SLOs that are tighter than those of other query types 
(\ie their percentile response times are closer to the values in their SLOs).
Since admitted queries share the same FIFO queue, it is possible that queries with looser SLOs cause
queries with tighter SLOs to be rejected in large numbers, especially at high QPS.
Figure~\ref{fig:query_starvation_example} illustrates the problem, which we reproduced by sending
traffic at a high rate to an experimental cluster running \liquid graph database
(\S\ref{sec:liquid}) with basic \bouncer.
We selected two types of queries, identified as \texttt{FAST} and \texttt{SLOW}, and generated
traffic from a set of millions queries sampled from production. 
Both queries types are configured with $SLO_{p50}$=18ms and $SLO_{p90}$=50ms, shown in dotted lines;
thus, the SLO is tighter for the \texttt{SLOW} queries than for the \texttt{FAST} ones.  
The figure reports the p50 and p90 response time estimates for both query types in a one-second
interval. 
We can see that the \texttt{FAST} queries make the \texttt{SLOW} queries ``starve'' as nearly 100\%
of the \texttt{SLOW} queries exceed the SLO and get rejected in the interval.
Therefore, \bouncer needs to alleviate query starvation, and we discuss two alternative 
strategies for that purpose next.

\subsection{Acceptance Allowance}
\label{sec:acceptance_allowance}

\begin{algorithm}[!tp]
  \small
  \SetAlgoNoLine
  \SetAlgoNoEnd

  \KwIn{$Q$: incoming query.}
  \KwData{
    $SW$: sliding window;
    $A \in [0.0, 1.0]$: acceptance allowance.
  }
  \KwOut{$decision$: Accept or Reject.}

  $type \gets Type(Q)$\;

  \tcp{Historical counts}
  $aqc \gets SW.GetAcceptedQueryCount(type)$\;
  $rqc \gets SW.GetQueryCount(type)$\;

  $decision \gets$ Reject\;
  \lIf{$rqc = 0$}{$decision \gets$ Accept}
  \Else{
    $AR \gets aqc / rqc$ \tcp*[l]{Acceptance ratio}
    \lIf{$AR < A$}{$decision \gets$ Accept}
  }

  \If{$decision =$ Reject}{
    $decision \gets Bouncer.CanAdmit(Q)$ \tcp*[l]{Ask the policy}
  }

  \If{$decision =$ Reject}{
    \tcp{On the spot}
    \lIf{$\mathtt{rand()} < A$}{$decision \gets$ Accept}
  }

  \If{$decision =$ Accept}{
    $SW.IncrementAcceptedQueryCount(type)$\;
  }
  $SW.IncrementQueryCount(type)$\;

  \caption{Acceptance allowance strategy. }
  \label{alg:acceptance_allowance}
\end{algorithm}

This strategy ensures that a small percentage of queries of each type is always admitted.
For that, it gives a little \emph{allowance} to each query type to get accepted and processed by the
system, hence its name. 
The strategy, shown in Algorithm~\ref{alg:acceptance_allowance}, operates on a sliding window $SW$
with duration $D$ and time step $\Delta$, where $D \gg \Delta$ (\eg $D$=1s and $\Delta$=10ms).
The sliding window tracks the number of accepted queries ($aqc$) and 
received\footnote{accepted and rejected} queries ($rqc$) per query type.
The parameter $A \in [0.0, 1.0]$ represents the \emph{acceptance allowance} and is expected to be
small (0.01--0.03).
Setting $A = 0.01$ means that we are willing to give ``free passes'' to up to $1\%$ of the queries
of each type over the span of the sliding window. 
Although admission decisions are made per query type, we use the same value of $A$ irrespective of
the type for the strategy to have few configuration parameters.

The call to \bouncer splits the strategy in two parts. 
The first part makes decisions based on the historical query counts ($aqc$ and $rqc$) and their
quotient (the acceptance ratio), while the second part overrides rejection decisions ``on the spot''
uniformly at random with probability equal to $A$.

Besides relieving query types from systemic service denial, 
by \emph{always} letting a few queries of the different types in,  
this strategy ensures that the processing time histograms \bouncer uses for admission decisions
get populated.

\subsection{Helping the Underserved}
\label{sec:help_underserved}

\begin{figure}[!tp]
  \centering
  \includegraphics[width=0.40\columnwidth]{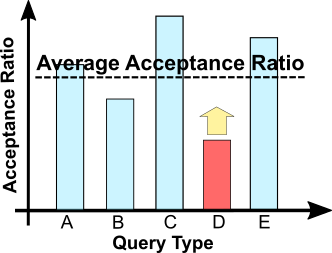}
  \vspace{-.40cm}
  \caption{Intent of helping the underserved.}
  \label{fig:help_underserved}
\end{figure}

\begin{algorithm}[!tp]
  \small
  \SetAlgoNoLine
  \SetAlgoNoEnd

  \KwIn{$Q$: incoming query.}
  \KwData{
    $SW$: sliding window;
    $QT$: set of query types;
    $\alpha \in (0.0, 1.0]$: scaling factor.
  }
  \KwOut{$decision$: Accept or Reject.}
  
  $decision \gets Bouncer.CanAdmit(Q)$ \tcp*[l]{Ask the policy}

  \If{$decision =$ Reject}{
    $type \gets Type(Q)$\;
    \tcp{Acceptance ratio for the query type}
    $AR \gets 
      \dfrac{SW.GetAcceptedQueryCount(type)}{\mathtt{max}(SW.GetQueryCount(type), 1)}$\;

    \tcp{Average acceptance ratio}
    {\footnotesize
    $AAR \gets 
      \sum_{t \in QT}\dfrac{SW.GetAcceptedQueryCount(t)}{\mathtt{max}(SW.GetQueryCount(t), 1)}$\;
    }
    $AAR \gets AAR / \mathtt{max}(|QT|, 1)$\; 

    \If{$AR < ARR$}{
      $x \gets (AAR - AR)/ARR$\; 
      $p \gets \alpha \cdot x/(1+x)$\;
      \lIf{\texttt{rand()} $<p$}{Accept}
    }   
  }
  \If{$decision =$ Accept}{
    $SW.IncrementAcceptedQueryCount(type)$\;
  }
  $SW.IncrementQueryCount(type)$\;

  \caption{``Helping the underserved'' strategy.
  }
  \label{alg:help_underserved}
\end{algorithm}

Rather than assigning a fixed allowance to the query types, 
this strategy is more dynamic and tries to help query types that have been rejected more than
others.
A query type is deemed that has been treated unfavorably when its \emph{acceptance ratio} (the
quotient between accepted and received queries) is lower than the \emph{average acceptance ratio}
across query types (see Figure~\ref{fig:help_underserved}).

The strategy is presented in Algorithm~\ref{alg:help_underserved}.
Like the previous one, it operates on a sliding window $SW$, with duration $D$ and time step
$\Delta$, which tracks the number of accepted and received queries per query type. 
A key step in this strategy is the calculation of $p$, which represents the probability of
overriding \bouncer's decision of rejecting a query $Q$ when the acceptance ratio for $Q$'s type is
lower than the average acceptance ratio ($AR < ARR$). 
We use a heuristic to calculate $p$. 
We do not make $p = (ARR-AR)/ARR$ because we would give unfavored query types excessive help 
(\ie if $AR$ is very small, then $p \approx 1$).
Instead, we use the sigmoid function $p = \alpha \cdot x/(1+|x|)$ to reduce and smooth ``the help''
given to query types, where $\alpha \in (0.0, 1.0]$ is a configurable scaling factor.

While this strategy may help \bouncer populate its histograms, it is not guaranteed to be
effective in doing so because its admission decisions are probabilistic based on the value of $p$.
For example, it is possible that a series of queries with a rarely-seen type may get unlucky and all
be rejected by both \bouncer and the strategy.

\section{Experimental Evaluation}
\label{sec:eval}

In this section, we evaluate \bouncer (\S\ref{sec:slo_policy}) and its variants with the starvation
avoidance strategies discussed in \S\ref{sec:query_starvation}.
We conduct two studies, one using simulation (\S\ref{sec:simulation_results}) and the other on
\liquid distributed graph database (\S\ref{sec:impeller_results}), that compare \bouncer with other policies. 
We first give an overview of \liquid and then describe the policies we evaluate \bouncer against. 
After presenting the studies, the section ends with a summary of the results
(\S\ref{sec:result_summary}).

\subsection{\liquid Graph Database}
\label{sec:liquid}
\begin{figure}[!tp]
\centering
\includegraphics[width=0.90\columnwidth]{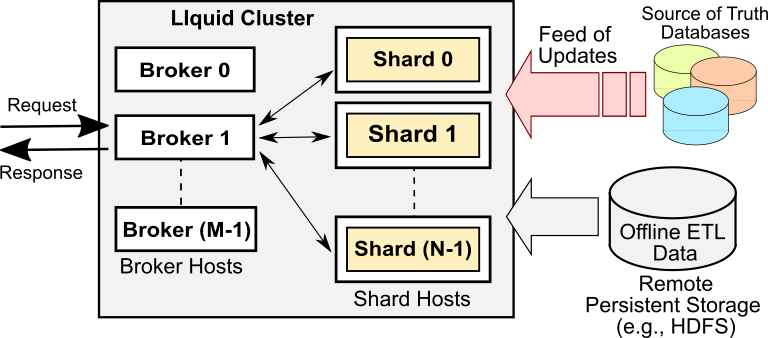}
\vspace{-.35cm}
\caption{
  \liquid cluster, unit of deployment.
}
\label{fig:liquid-arch}
\end{figure}
\liquid~\cite{liquid20a,liquid20b} is an in-memory distributed graph database built at LinkedIn to
answer online, interactive queries at high throughput and with low latency.
Among other data corpora, it serves LinkedIn’s Economic Graph~\cite{economicgraph} that includes
hundreds of millions of vertices and hundreds of billions of edges.
To store large graphs, \liquid is deployed on a cluster of machines organized in two tiers, 
\emph{brokers} and \emph{shards} (see Figure~\ref{fig:liquid-arch}). 
Our data centers host multiple \liquid clusters that act as replicas to serve large volumes of
traffic (several millions of queries per second) with high availability.

A \liquid cluster breaks up the graph into multiple data shards and assigns them to separate shard
hosts, which store and index the data in memory~\cite{carter19}.  
The shard hosts load data generated by offline jobs from remote persistent storage.
They also receive a continuous feed of updates (\eg via Kafka~\cite{kafka}) from source-of-truth
databases, and each shard keeps the updates belonging to its slice of the graph.

The broker hosts offer REST endpoints for clients to send query requests.
\liquid’s request processing is stateless as client requests are considered independent units. 
When a broker receives a query from a client, the broker sends sub-queries to the shard hosts to
fetch data from them.
Answering a query involves one or more communication rounds between the broker and the shards.
At the end of each round, the broker accumulates the shards’ responses and processes the sub-query
results before starting the next round.
Once the last communication round completes, the broker sends a response with the query result back
to the client.

Brokers and shards implement the admission control framework described in \S\ref{sec:slo_policy}.
They run a configurable number of query engine processes that cycle between obtaining an admitted
(sub-)query from the FIFO queue and processing it; 
these processes operate independently and in a lock-free manner. 
Brokers and shards also enforce expiration times for admitted queries. 
In our evaluation, shards, which do the heavy lifting in query processing, use the acceptance
fraction policy (\S\ref{sec:acceptrate}) governed by a utilization threshold, 
while brokers use \bouncer or one of the policies in \S\ref{sec:other_policies}.

\subsection{Other Admission Control Policies}
\label{sec:other_policies}

Here we describe several in-house policies available in \liquid and used in our
comparative evaluation.
They are built atop the framework described in \S\ref{sec:slo_policy}, but use
different criteria for decision making.
Unlike \bouncer, these policies are oblivious to query types. 

\subsubsection{\textbf{Maximum Queue Length (\mql) Policy}}
\label{sec:maxqueuelength}

It simply accepts an incoming query only if the FIFO queue's length is
less than a configurable length limit ($l < L_{limit}$)~\cite{iyer2001}.

\subsubsection{\textbf{Maximum Queue Wait Time (\mqwt) Policy}}
\label{sec:maxqueuewaittime}

It admits an incoming query $Q$ only if the estimate for $Q$'s mean queue wait
time is less than or equal to a configurable time limit
($ewt_{mean} \leq T_{limit}$). 
The mean queue wait time is estimated as: 
\begin{equation}
ewt_{mean} = l \times pt_{mavg}~/~P
\label{eqn:mean_wait_time}
\end{equation}
where 
$l$ is the FIFO queue's current length; 
$pt_{mavg}$ is the moving average of query processing times in a sliding window
of duration $D$ and time step $\Delta$, with $D \gg \Delta$; 
and
$P$ is the number of processes responsible for processing queries. 

\mqwt is an experimental policy we include to evaluate the use of the mean
queue wait time as a metric for per-query admission decision making,
rather than as an overload signal~\cite{zhou2018overload,breakwater2020}.
Its main configuration parameter is the maximum wait time, which it enforces
without differentiating query types.  
We discuss the effect of lifting this limitation in \S\ref{sec:mqwt_vs_slo}.
Unless stated otherwise, \mqwt's sliding window is configured with 
$D = \text{60s}$ and $\Delta = \text{1s}$.

\subsubsection{\textbf{Acceptance Fraction (\af) Policy}}
\label{sec:acceptrate}

It periodically computes the fraction of queries that the host should accept as:
$$f = min \left(1.0, \frac{\text{available processing capacity}}{\text{demanded processing capacity}}\right)$$
A host's \emph{available processing capacity} is the number of processing units the host has at its
disposal to process queries, 
while  
its \emph{demanded processing capacity} is the number of processing units it needs to fully serve
the incoming flux of queries.

This policy calculates the available processing capacity as $APC = MaxUtil \times |PU|$, 
where 
$MaxUtil \in (0.0, 1.0]$ is the maximum utilization threshold, 
and 
$|PU|$ is the number of processing units set aside for query processing.
$MaxUtil$ and $|PU|$ are configuration parameters and $APC$, once computed, remains fixed throughout
the system's operation.
Also, the policy periodically computes the demanded processing capacity, which generally varies over
time, as 
$dpc = qps_{mavg} \times pt_{mavg}$,
where 
$qps_{mavg}$ is the moving average of the incoming traffic rate in queries per second, 
and  
$pt_{mavg}$ is the moving average of the query processing times.
Both moving average values are obtained in a sliding window of duration $D$ and
time step $\Delta$, with $D \gg \Delta$.
Thus, the fraction of queries that should be accepted is given by:\footnote{
  $dpc = qps_{mavg} \times pt_{mavg}$ may be a small positive value or even zero. 
  We rely on standard floating-point arithmetic to handle these cases; \eg if $dcp=0.0$, then $f= min(1.0, inf)=1.0$.
}
$$f = min \left(1.0, \frac{MaxUtil \times |PU|}{qps_{mavg} \times pt_{mavg}}\right)$$
The policy then accepts queries with probability equal to the current value of $f$. 
If $f=1.0$ it accepts every query, but if $f<1.0$ it probabilistically rejects
$(1.0-f) \times 100\%$ of the queries (\ie the percentage exceeding the available capacity).

The number of processing units ($|PU|$) represents the level of parallelism for query processing on
a host. 
On shards, $|PU|$ is set as the number of CPU cores dedicated to process sub-queries from brokers,
whereas on brokers $|PU|$ is the number of processes responsible for handling queries from clients,
issuing sub-queries to and processing responses from the shards.

\af also estimates the mean queue wait time of every query using \EQ~\ref{eqn:mean_wait_time} with
$P=|PU|$, and rejects the queries expected to time out in the queue.
Unless stated otherwise, this policy is configured to update the demanded processing capacity
($dpc$) every second and with $D=\text{60s}$ and $\Delta=\text{1s}$ for the sliding window.

\subsection{Simulation Study}
\label{sec:simulation_results}

We compare the basic behavior of the policies listed in Table~\ref{tab:parameters}, using a
discrete event-driven simulator we wrote in Python~3.
The simulator implements the framework in Figure~\ref{fig:framework}.
It assumes a query engine with a fixed number of processes and gives the admitted queries to the
idle processes on a first-come, first-serve basis. 

We simulate a \liquid broker host (\S\ref{sec:liquid}) with 100 query engine processes, a number in
the same order of magnitude used in practice.
Table~\ref{tab:sim_parameters} lists the query types used in this study. 
Each type is given a fixed percentage among the generated queries (\ie its proportion in the query
mix), and its processing times follow a lognormal distribution, which approximates those of real
production queries.

\begin{table}[!tp]
\footnotesize
\centering
\caption{Query types used in the simulation study.}
\vspace{-.35cm}
\begin{tabular}{lcccccc}
  \begin{tabular}[c]{@{}c@{}} \textbf{Anonymized} \\ \textbf{Query Types}\end{tabular} & \begin{tabular}[c]{@{}c@{}} \textbf{Proportion in} \\ \textbf{Query Mix}\end{tabular} & $\mathbf{pt_{mean}}$ (ms) & $\mathbf{pt_{p50}}$ (ms) & $\mathbf{pt_{p90}}$ (ms)  \\ \hline
  \texttt{fast}         & 40\% & 1.16  & 0.38  & 2.70   \\ 
  \texttt{medium fast}  & 20\% & 2.53  & 2.22  & 4.27   \\ 
  \texttt{medium slow}  & 30\% & 12.13 & 7.40  & 26.44  \\ 
  \texttt{slow}         & 10\% & 20.05 & 12.51 & 44.26  \\ \hline
\end{tabular}
\label{tab:sim_parameters}
\end{table}

The traffic rate that fully utilizes the query engine is given by 
$QPS_{full\_load} = \frac{P}{pt_{wmean}}$,  
where $P$ is the number of query engine processes, and 
$pt_{wmean}$ is the weighted mean processing time of the query types with their proportions in the
query mix. 
We evaluate the policies with traffic rates ranging from $0.9 \times QPS_{full\_load}$ 
to $1.5 \times QPS_{full\_load}$. 
From the values in Table~\ref{tab:sim_parameters}, $pt_{wmean}$ 
is equal\footnote{$pt_{wmean} = 0.4 \cdot 1.16 + 0.2 \cdot 2.53 + 0.3 \cdot 12.13 + 0.1 \cdot 20.05 = 6.614\text{~ms}$}
to 6.614ms.
Then, with $P=100$, 
$QPS_{full\_load} = \frac{100}{6.614 \cdot 10^{-3}} \approx \text{15.1~kQPS}$, 
and 
the traffic rate range is 
$[\text{13.6~kQPS},\text{22.7~kQPS}]$.

The inter-arrival times for the queries are generated from an exponential distribution to
simulate traffic burstiness.
We subject the policies to the same incoming traffic and compare them on three dimensions: 
SLO violations, rejection ratio, and system utilization.
Each simulation run produces 1.5 million queries, lasting between one and two minutes of simulated time, 
and is preceded by a warm-up phase to avoid capturing cold start effects in our results. 
Table~\ref{tab:parameters} lists the parameters used for the policies in this study.

\begin{table}[!tp]
\footnotesize
\centering
\caption{Parameters for the admission control policies in the simulation study.}
\vspace{-.35cm}
\begin{tabular}{ll}
  \textbf{Policy}                & \textbf{Parameters} \\ \hline
  \bouncer                       & $SLO_{p50}$=18ms; $SLO_{p90}$=50ms   \\
  \bouncer + \aallow             & $A$=0.05 \\
  \bouncer + \htu                & $\alpha$=1.0 \\ \hline
  \mql                           & queue length limit = 400 \\ \hline 
  \mqwt                          & wait time limit = 15ms \\ \hline
  \af                            & utilization threshold=95\% \\ \hline
  \multicolumn{2}{p{0.95\linewidth}}{
    \bouncer is configured with the $SLO_{p50}$ and $SLO_{p90}$ values given above, even when
    supplemented with a starvation avoidance strategy.
  }
\end{tabular}
\label{tab:parameters}
\end{table}

\subsubsection{\textbf{Basic \bouncer Policy vs. The Other Policies}}
\label{sec:basic_slo_study}
\begin{figure}[!tp]
  \centering
  \includegraphics[width=0.70\columnwidth]{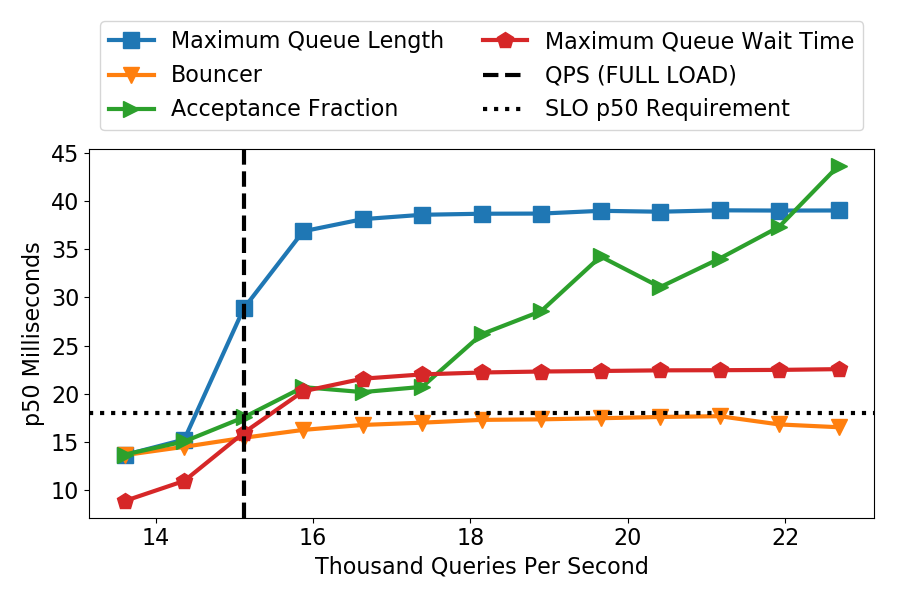}
  \vspace{-.45cm}
  \caption{
    Median response time ($rt_{p50}$) for \texttt{slow} queries.
    The latency SLO is the tightest for these queries (see Tables~\ref{tab:sim_parameters} and~\ref{tab:parameters}).
    \af policy's threshold utilization is 95\%.
  }
  \label{fig:sim_p50}
\end{figure}

Here we evaluate \bouncer's basic formulation (without starvation avoidance) against the policies in
\S\ref{sec:other_policies}. 
We observe that, when the traffic load exceeds $QPS_{full\_load}$, \bouncer keeps the serviced
queries within the latency SLO whereas the other policies do not.
Figure~\ref{fig:sim_p50} shows this result for the \texttt{slow} queries, for which the SLO is the
tightest; the other query types exhibit similar behavior. 

The median response time ($rt_{p50}$) for \mql, although above the SLO, plateaus at $\sim$40ms due
to the limit this policy imposes on queue length.
Similarly, \mqwt plateaus at a $rt_{p50}$ of $\sim$22ms because it limits the time the queries wait
in the queue; yet it exceeds the SLO since it does not take into account the percentile response
times of the query types. 
By contrast, $rt_{p50}$ for \af grows with the traffic rate, reaching the top, right corner of
Figure~\ref{fig:sim_p50}. 
The reason is that in our simulation this policy imposes no limits on queue length and queue wait
time.\footnote{
Note that this is not the case in \liquid, where \af rejects queries expected to time
out and enforces a maximum queue length to safeguard against very high response times.
}
Thus, when the system is overloaded, the queue length grows significantly and so does $rt_{p50}$.

\begin{figure}[!tp]
  \centering
  \includegraphics[width=0.70\columnwidth]{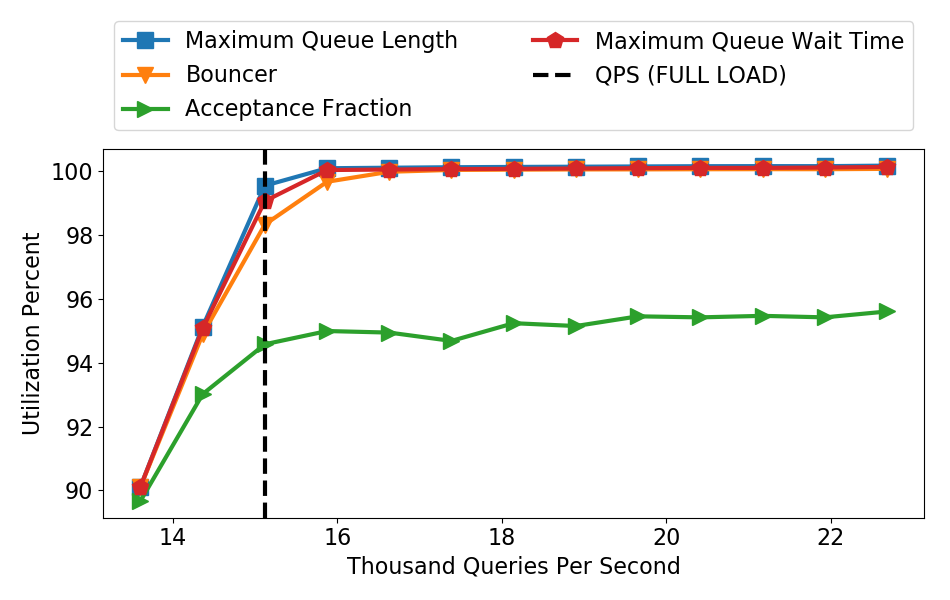}
  \vspace{-.45cm}
  \caption{
    System's utilization.
    \af policy's threshold utilization is 95\%.
  }
  \label{fig:sim_utilization}
\end{figure}

\begin{figure}[!tp]
  \centering
  \includegraphics[width=0.70\columnwidth]{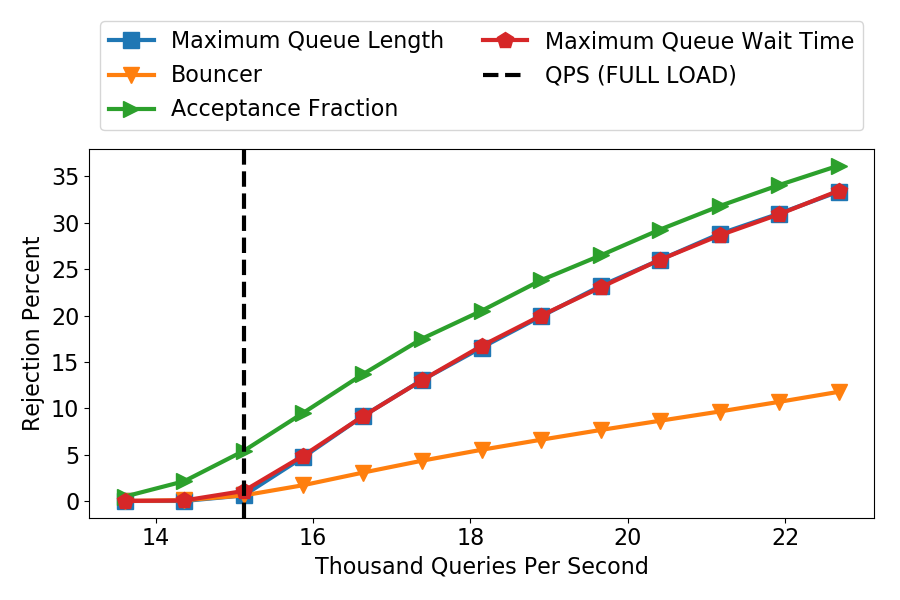}
  \vspace{-.45cm}
  \caption{
    Percentage of overall rejections.
  }
  \label{fig:sim_rejection}
\end{figure}

As shown in Figure~\ref{fig:sim_utilization}, all the policies but one are able to nearly reach
100\% utilization when the traffic approaches and exceeds $QPS_{full\_load}$; the exception is \af
that is limited by its utilization threshold of 95\%.
In addition, Figure~\ref{fig:sim_rejection} shows that as the QPS increases beyond
$QPS_{full\_load}$, the policies reject more queries to keep the system's load at bay. 
\bouncer reports the lowest percentage of rejections for the following reasons.
First, the queries it rejects the most are \texttt{slow} queries (see
Table~\ref{tab:sim_rejections_slo_and_strategies}). 
Second, compared to the other query types, the \texttt{slow} queries have the highest processing
time (closest to the latency SLO); thus, \bouncer needs to reject a lower number of queries of the
\texttt{slow} type to prevent the system from overloading.
The other policies report more rejections because they make no distinction between the types of
queries and treat them as equals. 
In particular, \af has the highest rejection percentage because it is bounded by a 95\% utilization
threshold while the other policies are not. 

Our simulation results show that \bouncer's basic formulation performs better than the other polices
during overload as it  keeps the queries within their SLOs and reports the least overall rejections,
while allowing the system reach nearly maximal utilization.

\subsubsection{\textbf{\bouncer policy with Starvation Avoidance Strategies}}
\label{sec:query_strarvation_study}

\begin{table*}[t]
\footnotesize
\centering
\caption{
Percentage of rejections reported by \bouncer with and without starvation avoidance strategies.
}
\vspace{-.35cm}

\begin{tabular}{llccccccccccccc}
 \hline
 &  & \multicolumn{13}{c}{\textbf{Factor of} $\mathbf{QPS_{full\_load}}$} \\ \cline{3-15} 
 \multirow{-2}{*}{\textbf{Policy (Strategy)}} & \multirow{-2}{*}{\textbf{Query Type}} & 0.9$\times$ & 0.95$\times$ & 1.0$\times$ & 1.05$\times$ & 1.1$\times$ & 1.15$\times$ & 1.2$\times$ & 1.25$\times$ & 1.3$\times$ & 1.35$\times$ & 1.4$\times$ & 1.45$\times$ & 1.5$\times$ \\ \hline
 & \texttt{fast}         & -0- & -0- & -0- & -0- & -0- & -0- & -0- & -0- & -0- & -0- & -0- & -0- & -0- \\
 & \texttt{medium fast}  & -0- & -0- & -0- & -0- & -0- & -0- & -0- & -0- & -0- & -0- & -0- & -0- & -0- \\
 & \texttt{medium slow}  & -0- & -0- & -0- & -0- & -0- & 0.00 & 0.00 & 0.01 & 0.05 & 0.23 & 0.82 & 2.29 & 4.86 \\
 & \texttt{slow}         & 0.01 & 0.53 & 5.02 & 15.89 & 29.27 & 41.84 & 53.63 & 64.37 & 74.18 & 82.88 & 90.37 & 95.68 & 98.46 \\
  \multirow{-5}{*}{\begin{tabular}[c]{@{}l@{}} \bouncer \\ (Basic \\ Formulation) \end{tabular}}
 & \texttt{ALL}          & 0.00 & 0.05 & 0.50 & 1.59 & 2.93 & 4.18 & 5.36 & 6.44 & 7.43 & 8.36 & 9.28 & 10.25 & 11.30 \\
 \hline
 & \texttt{fast}         & -0- & -0- & -0- & -0- & -0- & -0- & -0- & -0- & -0- & -0- & -0- & -0- & -0- \\
 & \texttt{medium fast}  & -0- & -0- & -0- & -0- & -0- & -0- & -0- & -0- & -0- & -0- & -0- & -0- & -0- \\
 & \texttt{medium slow}  & -0- & -0- & -0- & -0- & -0- & -0- & 0.02 & 0.07 & 0.36 & 1.29 & 3.45 & 6.86 & 10.83 \\
 & \texttt{slow}         & 0.01 & 0.53 & 4.97 & 15.98 & 29.31 & 41.86 & 53.58 & 64.24 & 73.56 & 80.97 & 85.63 & 87.58 & 88.12 \\
  \multirow{-5}{*}{\begin{tabular}[c]{@{}l@{}} \bouncer \\ (Acceptance \\ Allowance,\\ $A$=0.1)\end{tabular}}
 & \texttt{ALL}          & 0.00 & 0.05 & 0.50 & 1.60 & 2.93 & 4.19 & 5.36 & 6.45 & 7.46 & 8.48 & 9.60 & 10.82 & 12.06 \\
 \hline 
 & \texttt{fast}         & -0- & -0- & -0- & -0- & -0- & -0- & -0- & -0- & -0- & -0- & -0- & -0- & -0- \\
 & \texttt{medium fast}  & -0- & -0- & -0- & -0- & -0- & -0- & -0- & -0- & -0- & -0- & -0- & -0- & -0- \\
 & \texttt{medium slow}  & -0- & -0- & -0- & -0- & -0- & 0.04 & 0.25 & 1.35 & 4.06 & 7.96 & 12.20 & 16.29 & 20.36 \\
 & \texttt{slow}         & 0.01 & 0.53 & 5.03 & 15.99 & 29.34 & 41.89 & 53.21 & 61.93 & 67.08 & 69.26 & 70.21 & 70.84 & 71.37 \\
  \multirow{-5}{*}{\begin{tabular}[c]{@{}l@{}} \bouncer \\ (Helping the \\ Underserved,\\ $\alpha$=1.0)\end{tabular}}
 & \texttt{ALL}          & 0.00 & 0.05 & 0.50 & 1.60 & 2.94 & 4.20 & 5.40 & 6.60 & 7.93 & 9.31 & 10.68 & 11.97 & 13.25 \\
 \hline
 \multicolumn{15}{l}{
  Each cell reports the average of 5 simulation runs. -0- means absolute zero rejections.} \\
\end{tabular}
\label{tab:sim_rejections_slo_and_strategies}
\end{table*}

Table~\ref{tab:sim_rejections_slo_and_strategies} reports the rejection percentages for 
\bouncer with and without the starvation avoidance strategies at increasing traffic rates.  
Naturally, as the load grows, the overall rejections increase for the basic policy and the
strategies.  
\texttt{Slow} queries experience most rejections and are subject to starvation at high traffic
rates.
\texttt{Medium slow} queries also experience rejections but far less, whereas 
\texttt{fast} and \texttt{medium fast} queries are never rejected. 
This result is expected since the processing times of \texttt{slow} and \texttt{medium slow}
queries are the first and second closest to the latency SLO (see Tables~\ref{tab:sim_parameters}
and~\ref{tab:parameters}).

We observe that the \bouncer's basic formulation rejects more than 90\% of \texttt{slow} queries at traffic
rates $\geq 1.4 \times QPS_{full\_load}$.
By contrast, both strategies, \aallow (\S\ref{sec:acceptance_allowance}) and \htu
(\S\ref{sec:help_underserved}), keep rejections of \texttt{slow} queries below 90\%.

The \aallow strategy limits the rejections of \texttt{slow} queries by enforcing the configured
allowance of 10\%. 
Hence, as the traffic load increases, the strategy causes rejections to shift from \texttt{slow} to
\texttt{medium slow} queries (up to $\sim$11\%), when compared to basic \bouncer. 
The reason is that the \texttt{slow} queries the strategy lets in take the room of other, less
costly query types, and the \texttt{medium slow} queries, being the second most costly, are next in
line to experience the effects.  
The \htu strategy behaves similarly, but only rejects up to $\sim$71\% of \texttt{slow} queries. 
The reason is that with $\alpha=1.0$, the probability of overriding \bouncer's rejections can get
close to 0.5.  
Thus, by rejecting less \texttt{slow} queries, the strategy forces more rejections of
\texttt{medium slow} queries (up to $\sim$20\%).

Compared to \bouncer's basic formulation, the starvation avoidance strategies report a small rise in
overall rejection percentage (up to $\sim$1\% and $\sim$2\% increase for \aallow and \htu,
respectively).
The strategies are expected to increase overall rejections because to make room for some additional
\texttt{slow} queries, a larger number of less expensive queries need to be rejected.
Yet, the rise is very modest because only \texttt{slow} and \texttt{medium slow} queries suffer
rejections, but the much cheaper \texttt{fast} and \texttt{medium fast} queries never do. 
This also means that \bouncer with starvation avoidance still offers fewer rejections when compared
to \mql, \mqwt and \af (see Figure~\ref{fig:sim_rejection}).

\begin{figure}[!tp]
  \centering
  \includegraphics[width=0.65\columnwidth]{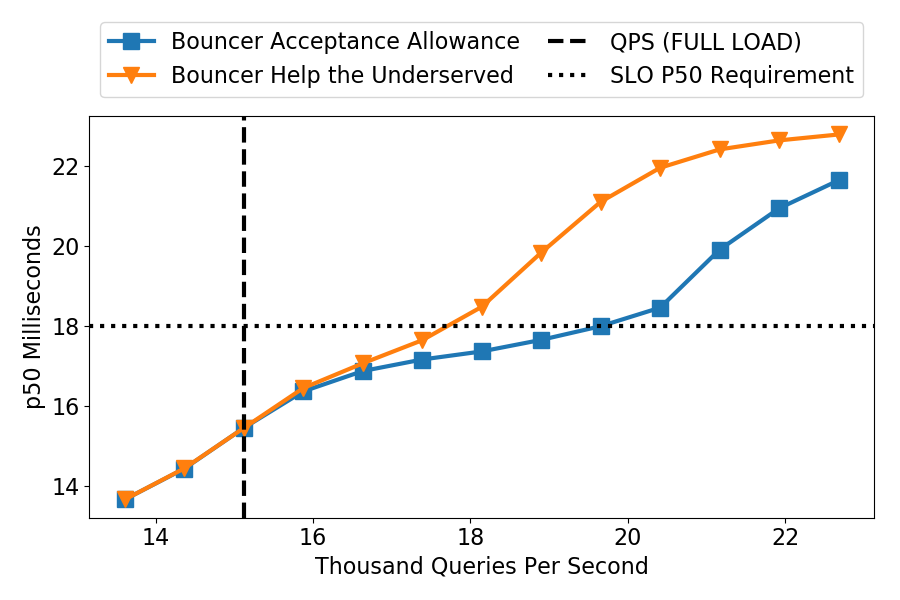}
  \vspace{-.45cm}
  \caption{
    Median response time ($rt_{p50}$) for \texttt{slow} queries.
  }
  \label{fig:sim_p50_improved}
\end{figure}

Also as expected, the strategies cause the response times to exceed the latency SLOs since to avoid
query starvation they must accept some requests that \bouncer's basic formulation would otherwise
reject. 
Our results indicate that \aallow is more advantageous than \htu in this regard.  
Figure~\ref{fig:sim_p50_improved} shows that \aallow lets the \texttt{slow} queries reach a higher
QPS without violating the $SLO_{p50}$ requirement, when compared to \htu.
Moreover, \aallow reports lower median response times for the \texttt{slow} queries at high traffic
rates.  
\Aallow's advantage  over \htu stems from how they detect query starvation. 
\Aallow considers that a query type is starving when its acceptance ratio ($AR$),
\emph{independently} from the other query types, falls under the given allowance ($A$).   
\Htu, on the other hand, considers that queries of a given type are starving when they are treated
unfavorably \emph{relative to} the other query types.
Our results suggest that relative unfavorable treatment among query types tend to be observed at
traffic rates lower than those at which per-query-type low acceptance ratios are. 

Like \bouncer's basic formulation, both strategies allow the system to reach close to 100\%
utilization, which is expected as they accept requests that plain \bouncer would reject.

\subsubsection{\textbf{Influence of Parameters on Query Starvation Avoidance}}
\label{sec:starvation_parameter_study}

\begin{table*}[t]
\footnotesize`
\centering
\caption{
  Percentage of rejections of \bouncer with the \aallow strategy for increasing values of parameter
  $A$ and traffic rate equal to $1.5 \times QPS_{full\_load}$.
}
\vspace{-.35cm}
\begin{tabular}{lcccccccccccc}
  \hline
  \multirow{4}{*}{\textbf{Query Type}} 
    & \multicolumn{11}{c}{$\mathbf{A}$} \\
    & \multicolumn{11}{c}{$[(1-A)\times100\%]$: Maximum rejection percentage enforced} \\ 
    \cline{2-13} 
    & \textbf{0.01} & \textbf{0.02} & \textbf{0.03} & \textbf{0.04} & \textbf{0.05} & \textbf{0.06} & \textbf{0.07} 
      & \textbf{0.08} & \textbf{0.09} & \textbf{0.1} & \textbf{0.2} & \textbf{0.3}  \\ 
    & [99\%] & [98\%] & [97\%] & [96\%] &  [95\%]  & [94\%] & [93\%] & [92\%] & [91\%] & [90\%] & [80\%] & [70\%] \\
  \hline
  \texttt{fast}        & -0- & -0- & -0- & -0- & -0- & -0- & -0- & -0- & -0- & -0- & -0- & -0-\\
  \texttt{medium fast} & -0- & -0- & -0- & -0- & -0- & -0- & -0- & -0- & -0- & -0- & -0- & -0-\\
  \texttt{medium slow} & 5.56 & 6.08 & 6.64 & 7.24 & 7.72 & 8.38 & 9.04 & 9.57 & 9.96 & 10.74 & 16.49 & 22.26\\
  \texttt{slow}        & 97.21 & 96.23 & 95.25 & 94.30 & 93.26 & 92.19 & 91.20 & 90.17 & 89.16 & 88.13 & 77.48 & 67.26\\
  \texttt{ALL}         & 11.39 & 11.45 & 11.52 & 11.60 & 11.64 & 11.73 & 11.83 & 11.89 & 11.91 & 12.03 & 12.70 & 13.40\\ 
  \hline
 \multicolumn{13}{l}{
  Cells report the average of 5 simulation runs. -0- means absolute zero rejections.} \\
\end{tabular}
\label{tab:rejections_varying_allowance}
\end{table*}

\begin{table*}[t]
\footnotesize`
\centering
\caption{
  Percentage of rejections of \bouncer with the \htu strategy for increasing values of parameter
  $\alpha$ and traffic rate equal to $1.5 \times QPS_{full\_load}$.
}
\vspace{-.35cm}
\begin{tabular}{lcccccccccc}
 \hline  
 \multirow{4}{*}{\textbf{Query Type}} 
    & \multicolumn{10}{c}{$\mathbf{\alpha}$} \\ 
    & \multicolumn{10}{c}{$[p_{max} \times 100\%]$: Maximum probability of acceptance by overriding rejection decision} \\ \cline{2-11} 
    & \textbf{0.1} & \textbf{0.2} & \textbf{0.3} & \textbf{0.4} & \textbf{0.5} & \textbf{0.6} & \textbf{0.7} & \textbf{0.8} & \textbf{0.9} & \textbf{1.0} \\ 
    & [5\%] & [10\%] & [15\%] & [20\%] & [25\%] & [30\%] & [35\%] & [40\%] & [45\%] & [50\%] \\ \hline
 \texttt{fast}        & -0- & -0- & -0- & -0- & -0- & -0- & -0- & -0- & -0- & -0- \\
 \texttt{medium fast} & -0- & -0- & -0- & -0- & -0- & -0- & -0- & -0- & -0- & -0- \\
 \texttt{medium slow} & 7.07 & 9.01 & 10.98 & 12.60 & 14.19 & 15.98 & 16.97 & 17.99 & 19.10 & 20.41 \\
 \texttt{slow}        & 94.74 & 91.32 & 88.11 & 84.81 & 82.38 & 79.47 & 77.10 & 75.01 & 72.98 & 71.15 \\
 \texttt{ALL}         & 11.59 & 11.83 & 12.11 & 12.26 & 12.50 & 12.74 & 12.80 & 12.90 & 13.03 & 13.24 \\ 
 \hline
 \multicolumn{11}{l}{
   From Algorithm~\ref{alg:help_underserved}, $p_{max}=\alpha\times1/2$. Cells report the average of 5 simulation runs. -0- means absolute zero rejections.
  } \\
\end{tabular}
\label{tab:rejections_varying_alpha}
\end{table*}

Here we study how $A$ and $\alpha$ influence the \aallow strategy and the \htu
strategy, respectively. 
We simulate an overload scenario with $QPS=1.5 \times QPS_{full\_load}$ and vary $A$ in the range
$[0.01, 0.3]$ and $\alpha$ in $[0.1, 1.0]$.   

Predictably, Table~\ref{tab:rejections_varying_allowance} shows that as $A$ increases, \aallow
rejects fewer \texttt{slow} queries, and the rejections shift from \texttt{slow} queries,
which are the most expensive, to  \texttt{medium slow} queries, the second most expensive. 
The rejection counts do not exceed the maximum rejection percentage enforced by the strategy.

Similarly, in the case of \htu, Table~\ref{tab:rejections_varying_alpha} shows that by increasing
$\alpha$, \texttt{slow} queries experience less rejections because the chances of \bouncer's
rejection decisions being overridden ($p_{max}$) grow.
Rejections also shift from \texttt{slow} to \texttt{medium slow} queries.
However, this strategy rejects more than $(1-p_{max})\times100\%$ of \texttt{slow} queries in most
cases; \eg with $\alpha=0.6$ and $p_{max}=0.3$, it rejects $\sim$79.5\% of \texttt{slow} queries
instead of 70\%, and with $\alpha=1.0$ and $p_{max}=0.5$, it rejects $\sim$71\% instead of 50\%. 
The reason is not only the probabilistic nature of \htu, but the fact that the probability $p$ of
overriding rejections is not always close to its maximum because it depends on two values that vary
over time -- the query type's acceptance ratio ($AR$) and the overall average acceptance ratio
($ARR$).
By contrast, a given $A$ in \aallow \emph{directly} means having a rejection percentage no larger
than $(1-A)\times100\%$, which makes \aallow more intuitive to use than \htu.

Tables~\ref{tab:rejections_varying_allowance} and~\ref{tab:rejections_varying_alpha} report similar
percentages of overall rejections (from 11.4\% to 13.4\%) for both strategies in the considered
ranges of $A$ and $\alpha$. 
The overall rejections also increase slightly for both strategies with their corresponding
parameters. 
As explained in \S\ref{sec:query_strarvation_study}, this is expected because when a strategy lets a
number of \texttt{slow} queries in, they occupy the room of a larger number of cheaper queries (\eg
\texttt{medium slow}) that need to be rejected. 

\begin{figure}[!tbp]
  \centering
  \includegraphics[width=0.5\columnwidth]{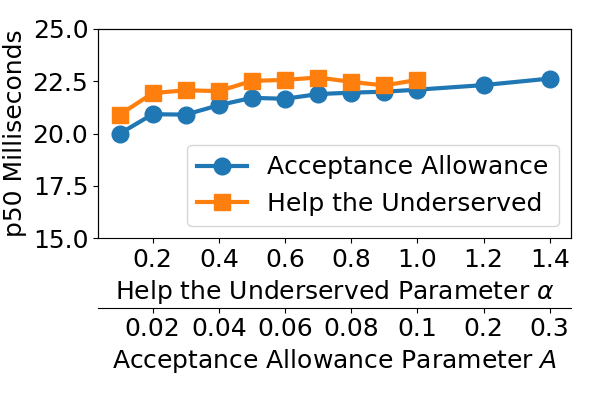}
  \vspace{-.45cm}
  \caption{
    Median response time ($rt_{p50}$) for \texttt{slow} queries with different parameter
    values for \bouncer's starvation avoidance strategies.
  }
  \label{fig:qs_p50}
\end{figure}

We also observe that varying $A$ and $\alpha$ has no significant impact on the query response times.
For example, Figure~\ref{fig:qs_p50} shows that $rt_{p50}$ for \texttt{slow} queries grows very
slowly with $A$ and $\alpha$ (less than 10\% increase).
Both strategies report similar $rt_{p50}$ (above 20ms), and as in Figure~\ref{fig:sim_p50_improved},
$rt_{p50}$ exceeds the $SLO_{p50}$ requirement (18ms) because the strategies accept additional
requests that \bouncer's basic formulation would not. 
Finally, for the considered values of $A$ and $\alpha$, both strategies let the system nearly reach
100\% utilization.

\subsection{Study on a Real System}
\label{sec:impeller_results}

Previously, we evaluated the admission control policies on a simulator assuming an ideal parallel
query processing engine. 
Now we evaluate them on \liquid graph database (\S\ref{sec:liquid}).
Our test platform is an experimental cluster with 16 shards and 12 brokers, serving a large portion
of LinkedIn's Economic Graph. 
Shards have 1.5TB of RAM and brokers 256GB, and each host is equipped with 
two Intel\textsuperscript{\textregistered} Xeon\textsuperscript{\textregistered} Platinum 8171M CPUs (52 cores) 
and a 25-Gbps NIC.

We compare \bouncer, supplemented with the two starvation avoidance strategies presented in
\S\ref{sec:query_starvation}, against the policies in \S\ref{sec:other_policies}.
We vary the policy on the brokers, while the shards always run \af (\S\ref{sec:acceptrate}).
The brokers are configured with the same policy for each test run.
Since the brokers are the queries's entry point to the \liquid cluster (see
Figure~\ref{fig:liquid-arch}), this setup enables the policies to reject queries early, while
letting \af guard against excessive CPU utilization on the shards, where CPU is the limiting
resource.

We send traffic to the \liquid cluster with a load generator program we built by modifying
\wrk~\cite{wrk2}, an open-source HTTP benchmarking tool capable of producing significant load. 
Our load generator sends 
HTTPS requests at an average rate given by the user, and 
emulates traffic burstiness with inter-departure times following an
exponential distribution.
It draws queries from one or more \emph{query sets}, each containing queries of a specific type, and
generates traffic according to a \emph{query mix}, which indicates the proportions per query type. 
The query sets and query mix are provided in input files.

The queries used in this study were obtained from online production traffic.
The query mix is given in the following list of \emph{query types} and their \emph{proportions}:
\{
\texttt{"\textbf{QT1}"}:  11.56\%, 
\texttt{"\textbf{QT2}"}:     0.04\%,
\texttt{"\textbf{QT3}"}:     0.04\%,
\texttt{"\textbf{QT4}"}:     2.34\%,
\texttt{"\textbf{QT5}"}:    13.44\%,
\texttt{"\textbf{QT6}"}:    13.44\%,
\texttt{"\textbf{QT7}"}:     0.42\%,
\texttt{"\textbf{QT8}"}:     0.09\%,
\texttt{"\textbf{QT9}"}:    26.35\%,
\texttt{"\textbf{QT10}"}:    4.49\%, 
\texttt{"\textbf{QT11}"}:   27.80\%
\}.
While anonymized, the query types are sorted by cost in ascending order.
We sampled the traffic of a \liquid cluster for 2 hours of daily peak load, for
2 weekdays to create the query sets in the mix, totalling 5.5 million queries.
Our selection of query types, out of several dozens, captures those with larger representation
as well as their diversity in processing time. 

We evaluate the policies at 36K, 72K, 108K, 144K, and 180K QPS.
Shards report high CPU utilization at rates $\geq$108K QPS.
Each test run lasts 10 minutes, in which our load generator issues queries to the cluster at one of
these traffic rates and with the load evenly divided among the brokers. 
We give queries generous expiration times to ensure they do not time out.
To avoid cold start effects, we warm up the cluster before each run by
sending traffic to it at the expected rate for one (extra) minute.   

Like in the simulation study, we configure \bouncer with $SLO_{p50}=\text{18ms}$ and
$SLO_{p90}=\text{50ms}$ for both starvation avoidance strategies; these values are similar to our
latency objectives in production for most queries in the mix.
The \aallow strategy uses $A=0.05$ and the \htu strategy $\alpha=1.0$.
\af is configured with a maximum utilization of 80\%, and \mqwt is given 12ms as its wait time
limit.
In \liquid not only \mql, but the other policies too can enforce a limit on the queue's length to
safeguard against its unbounded growth. 
We set the maximum queue length ($L_{limit}$) to 800 for all the policies.

\subsubsection*{\textbf{Results}}

\begin{figure}[!tp]
  \centering
  \includegraphics[width=0.50\columnwidth]{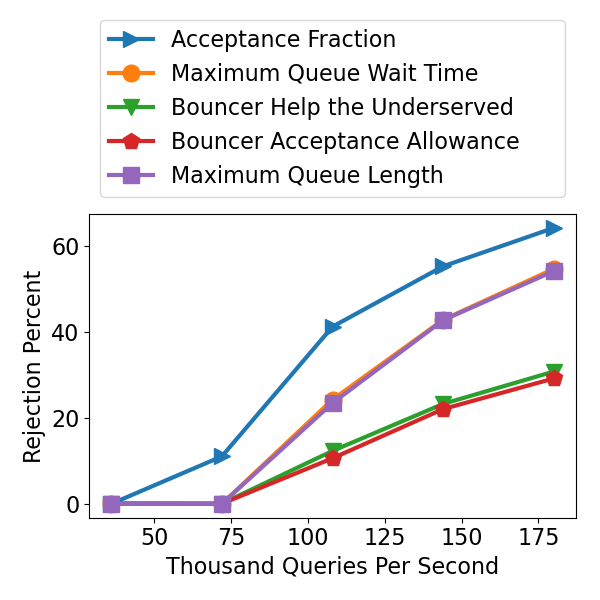}
  \vspace{-.45cm}
  \caption{
    Percentage of overall rejections on LIquid graph database (real system).
  }
  \label{fig:real_rr}
\end{figure}

Figure~\ref{fig:real_rr} shows that the overall rejections increase with the traffic load.
As expected, \bouncer's starvation-avoiding variants record similar rejection counts.
More importantly, they report lower percentages of overall rejections (between $\sim$15\% and
$\sim$30\% less) when compared to the other policies.  
With a similar result from simulation (Figure~\ref{fig:sim_rejection}), we confirm our observation
that \bouncer rejects fewer requests because it targets query types whose processing times are
higher and closer to the latency SLO, such as \texttt{QT11}.
By contrast, the other policies, oblivious to query types and their individual costs, refuse service
to cheap and expensive queries alike. 
Since rejecting a number of less expensive queries has a similar effect as rejecting a smaller
number of costlier ones, these policies report larger rejection counts. 
Figure~\ref{fig:real_rr} reports similar rejection percentages for the policies \mql and \mqwt, and
\af yields the highest rejection rates because its maximum utilization was conservatively set to
80\%.
We inspected the system's logs and confirmed that in the case of \mql, \mqwt and \bouncer, the
brokers, not the shards, produced the vast majority of rejections, and in the case of \mqwt, \af and
\bouncer, the brokers did not reach the queue length limit ($L_{limit}$).

\begin{figure}[!tp]
	\centering
	\begin{subfigure}[b]{1.0\columnwidth}
		\centering
		\includegraphics[width=0.5\columnwidth]{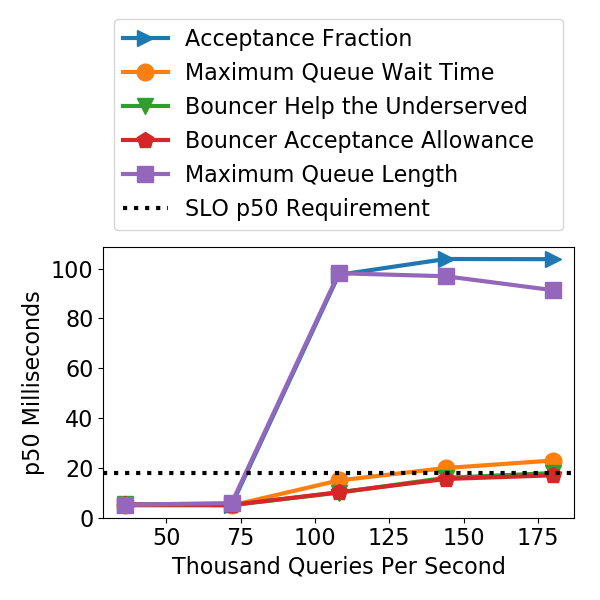}
		\vspace{-.35cm}
		\caption{
      50th-percentile response time ($rt_{p50}$).
		}
		\label{fig:real_sw_p50}
	\end{subfigure}

	\begin{subfigure}[b]{1.0\columnwidth}
		\centering
		\includegraphics[width=0.5\columnwidth]{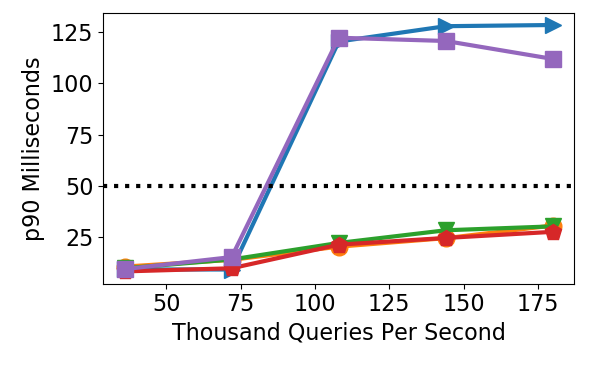}
		\vspace{-.35cm}
		\caption{
      90th-percentile response time ($rt_{p90}$).
   }
		\label{fig:real_sw_p90}
	\end{subfigure}

	\vspace{-.35cm}
	\caption{
   Response times for serviced \texttt{QT11} queries on LIquid graph database (real system).
	}
	\label{fig:real_sw}
\end{figure}

We now focus on the SLO violations for the \texttt{QT11} queries as they exhibit the largest
processing time (\ie the SLO is the tightest for them) and have the largest representation in the
mix.
Like in our simulation,  
Figure~\ref{fig:real_sw} shows that at meeting latency
requirements, \bouncer, with both starvation avoidance strategies, and 
\mqwt perform better than \mql and \af.
\bouncer and \mqwt maintain the response times of the serviced \texttt{Q11} queries close to
$SLO_{p50}$ and comfortably under $SLO_{p90}$.
Instead, \mql and \af let them exceed $SLO_{p50}$ ($>4\times$) and $SLO_{p90}$
($>2\times$) at traffic rates 
$\geq$108K QPS.

In this experiment, \bouncer rejects \texttt{QT11} queries more than any other; thus, they
are the only queries that suffer from starvation. 
Figure~\ref{fig:real_sw_p50} shows that at 144K and 180K QPS, \htu permits \texttt{QT11} queries to
slightly exceed $SLO_{p50}$ because it takes in some extra queries of this type to make up for
having treated them unfavorably.
By contrast, \aallow (with $A=0.05$) never acts since \texttt{QT11}'s rejections remain below 95\%;
hence, the median response time ($rt_{p50}$) for these queries stays under $SLO_{p50}$.

\begin{figure}[!tp]
  \centering
  \includegraphics[width=0.5\columnwidth]{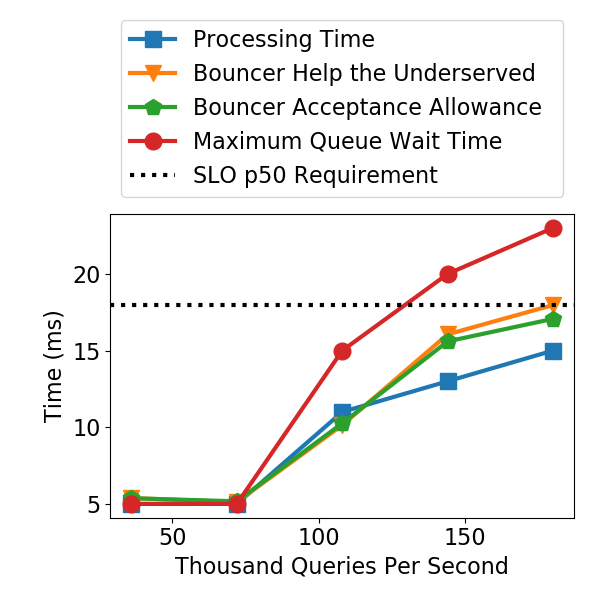}
  \vspace{-.45cm}
  \caption{
    Median processing time ($pt_{p50}$) for serviced \texttt{QT11} queries vs. 
    their median response time ($rt_{p50}$) under the \mqwt policy and 
    \bouncer with starvation avoidance, on LIquid graph database (real system).
  }
  \label{fig:real_processing_time}
\end{figure}

Figure~\ref{fig:real_sw_p50} also shows that under \mqwt, the \texttt{QT11} queries
exceed $SLO_{p50}$ at 144K and 180K QPS. 
We got this result despite our efforts to tune the policy's wait time limit for reducing rejection
counts and SLO violations, and setting it to 12ms ($< SLO_{p50}$=18ms). 
The reason is that the query processing time, observed by the brokers in the cluster, increases with
the traffic load.
As shown in Figure~\ref{fig:real_processing_time}, \texttt{QT11}'s median processing time
($pt_{p50}$) rises with the QPS and reaches $\sim$15ms at 180K QPS, just 3ms below $SLO_{p50}$.   
This real system's behavior differs from what we assumed in our simulation study.
Unlike an ideal parallel query engine, the \liquid cluster's processing tier, depicted in
Figure~\ref{fig:liquid-arch}, comprises shard hosts with their own FIFO queue\footnote{among
other queues in the NIC, OS, and implementation software.} and subject to queueing effects too. 
Figure~\ref{fig:real_processing_time} also reports that \mqwt lets \texttt{QT11}'s $rt_{p50}$ depart
from $pt_{p50}$.  
By contrast, under \bouncer, which considers both queue wait time and percentile processing times, 
\texttt{QT11}'s $rt_{p50}$ stays below or barely exceeds $SLO_{p50}$ and tracks
much more closely $pt_{p50}$. 
Thus, limiting the queue wait time is not enough to ensure that serviced queries meet the latency
SLO on the real system. 

Finally, our implementation of \bouncer reports a small overhead (mean=18$\mu$s,
p50=15$\mu$s, and p99=87$\mu$s) for millisecond-scale response times.

\subsection{\bouncer vs. \mqwt with parameter settings per query type}
\label{sec:mqwt_vs_slo}

\begin{figure}[!tbp]
	\centering
	\begin{subfigure}[b]{1.0\columnwidth}
		\centering
		\includegraphics[width=0.5\columnwidth]{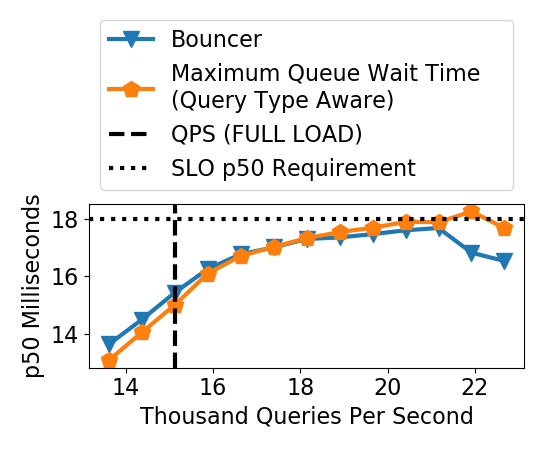}
		\vspace{-.35cm}
		\caption{
      Median response time ($rt_{p50}$) for \texttt{slow} queries 
      from Table~\ref{tab:sim_parameters}.
		}
		\label{fig:p50rt_compare_slo_mqwt}
	\end{subfigure}

	\begin{subfigure}[b]{1.0\columnwidth}
		\centering
		\includegraphics[width=0.5\columnwidth]{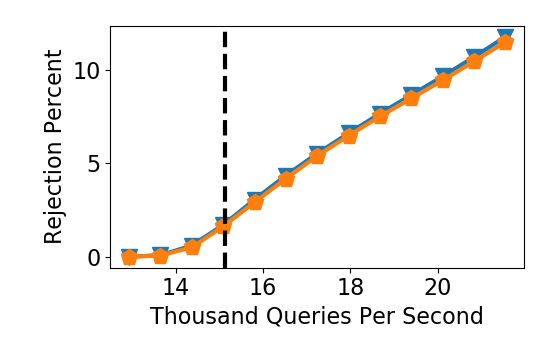}
		\vspace{-.35cm}
		\caption{Percentage of overall rejections.}
		\label{fig:rejections_compare_slo_mqwt}
	\end{subfigure}

	\vspace{-.35cm}
	\caption{
		\bouncer (without starvation avoidance) vs. 
    \mqwt with different wait time limits per query type. 
	}
	\label{fig:compare_slo_mqwt}
\end{figure}

As mentioned in \S\ref{sec:maxqueuewaittime}, the \mqwt policy evaluated in this paper is
experimental and receives a single parameter value -- the \emph{wait time limit}, which is enforced
without differentiating query types. 
Given this limitation of implementation and the relation between queue wait
time and response time (\EQ~\ref{eqn:response_time}), we ask ourselves: 
\emph{How does \bouncer, which allows individual SLOs per query type,
compare to \mqwt when wait time limits are also assigned per query type?}
We answer the question via simulation, and Figure~\ref{fig:compare_slo_mqwt} shows the median
response times for \texttt{slow} queries and overall rejections.

We observe that with properly chosen wait time limits per query type, \mqwt can
match \bouncer's behavior in terms of serviced queries meeting latency SLOs and
overall rejections. 
But, finding the right values is a time-consuming task of 
experimental tuning, 
and needs to be repeated for different workloads and their
causal variations.
Besides, the selected values may still be unreliable
because, as shown in Figure~\ref{fig:real_processing_time}, query processing times in real systems
may increase with the load. 
By contrast, \bouncer requires no tuning since it explicitly receives the
latency requirements in the SLOs and by design ensures that serviced queries
meet these requirements, even when the system is overloaded.  
Hence, from a practical standpoint \bouncer is more advantageous than \mqwt.

\subsection{Summary of Experimental Results}
\label{sec:result_summary}

Our results show, in simulation and on a real system, that under \bouncer serviced queries meet or,
when starvation avoidance is in use, track closely their latency SLOs.
The reason is that \bouncer uses percentile response time estimates obtained by combining
measurement-based approximations of queue wait time and percentile query processing times
(see Equations~\ref{eqn:ewt}, \ref{eqn:ert_p50}, and \ref{eqn:ert_p90}).
Further, despite their inherent accuracy loss, the adopted estimates have proven to enable effective
admission control decisions.

Our results also indicate that under \mqwt serviced queries are susceptible to violate SLOs
expressed in percentile response times ($SLO_{p50}$ and $SLO_{p90}$). 
The reason is that this policy makes decisions based on queue wait time estimates calculated using
the moving average processing time ($pt_{mavg}$) across queries of different types and cost (see
\EQ\ref{eqn:mean_wait_time}), and $pt_{mavg}$ is very different from the $pt_{p50}$ and $pt_{p90}$
values for individual query types.
Moreover, \mqwt can be configured with carefully chosen wait time limits to have admitted queries
meet their latency SLOs. 
But, it is generally impractical due to the laborious tuning involved, and it may even be
ineffective since query processing times in real database systems tend to increase with the load
(see Figure~\ref{fig:real_processing_time}).

\mql and \af, in contrast, do not enforce latency SLOs because that is not their design goal. 
The former only considers the number of queries in the queue (a coarse-grained queuing metric), and
the latter rejects queries to prevent the system from exceeding a utilization threshold.

We have also shown that both strategies, \aallow (\S\ref{sec:acceptance_allowance}) and \htu
(\S\ref{sec:help_underserved}), are able to avoid query starvation at the expense of  
\begin{inparaenum}[1)]
\item
a very modest increase in overall rejections (still lower than the other policies' rejection
counts), and 
\item
causing serviced queries to exceed the SLOs at high loads. 
\end{inparaenum}
Queries exceed the SLOs by a fair margin in simulation (Figure~\ref{fig:sim_p50_improved}), but
track them very closely in our experiments on \liquid (Figure~\ref{fig:real_sw}); this suggests that
the starvation avoidance strategies may cause acceptable counts of SLO violations in practice. 
Moreover, we find \aallow to be more advantageous than \htu because the allowance parameter $A$ of
the former is intuitive whereas the parameter $\alpha$ of the latter is not.  
Plus, in simulation \aallow activated at higher traffic rates compared to \htu.

Compared to \mql, \mqwt and \af, \bouncer reported the least overall rejections (between 15\% and
30\% less) by targeting the query types with the highest cost and tightest latency SLO. 
It also incurred small overhead (mean=18$\mu$s) for millisecond-scale queries and, with the given
latency SLOs, let the system reach nearly maximal utilization.

In our experiments on the \liquid cluster (\S\ref{sec:impeller_results}), 
we ran \bouncer on the brokers and \af on the shards. 
This pairing is reasonable because 
\af guards against excessive CPU usage on shards, where CPU is the limiting resource, while
\bouncer guards against violations of latency SLOs on the brokers, offering early rejections to
client requests. 
Thus, \bouncer is a viable complement to \af and other policies that protect the system from
exceeding its capacity.

\section{Related Work} 
\label{sec:related_work}

Admission control has been applied to 
networking~\cite{jamin1997,nam2008,yang2016joint},
operating systems~\cite{mittos2017},
web services~\cite{chen2002session,welsh03,quorum2005,zhou2018overload},
complex event processing over data streams~\cite{hspice2020,zhao2020,darling2022},
and
databases~\cite{elnikety2004,qcop2010,activesla2011,zhang2014,sap_hana_rm2019,sap_hana_ac,db2_wlm,oracledb_rm,sqlserver_rg}.
Here we discuss prior work closely related to \bouncer. 

Admission control techniques can be divided in two groups: 
1)~those that distinguish between request types
(\eg~\cite{elnikety2004,quorum2005,qcop2010,activesla2011}),
and
2)~those that do not 
(\eg~\cite{weikum1992,schroeder2006b,sert2006,selfstar2009}).
Like \bouncer, Gatekeeper~\cite{elnikety2004} belongs to the first group and implements a
measurement-based approach, but with different goals and response time estimates. 
Gatekeeper lets the system serve a sustained throughput without exceeding its capacity, and uses
moving averages to estimate mean response times.
\bouncer instead ensures that serviced requests meet or track closely their latency SLOs and
maintains histograms to estimate percentile response times.
While \bouncer strives for early rejections, Gatekeeper does not as it lets the requests time out in
the scheduling queue.
Gatekeeper implements a shortest-job-first (SJF) scheduler with an aging mechanism~\cite{alpha1998}
to reduce the overall average response time and prevent longer-running queries from starving. 
We also supplement \bouncer with starvation avoidance strategies that, rather than deferring query
executions, make decisions on the spot at query arrival.
\liquid currently processes queries in FIFO order and evaluating other scheduling disciplines is
left as future work.

Quorum~\cite{quorum2005} also differentiates request types and implements a non-invasive,
measurement-based approach to quality of service for Internet services. 
Quorum and \bouncer perform admission control at the system's entrance point, but for Quorum the
system is an Internet site whereas for \bouncer a two-tier database. 
Quorum ensures that serviced requests of different types meet their
response time guarantees, expressed as averages or percentiles. 
It selectively drops requests after being dequeued; 
by contrast, \bouncer rejects requests before being enqueued. 
Quorum prevents request types from starving by ensuring they get a large enough fraction
of capacity to guarantee their required minimum throughput at all times.  
Contrarily, \bouncer does not control the capacity given to query types and applies another kind
of strategies to avoid query starvation.

Q-Cop~\cite{qcop2010}, like \bouncer, focuses on admission control, but seeks to minimize the number
of request timeouts.
Q-Cop builds a linear regression model from offline experiments, and uses the model to predict the
processing time of a newly arriving query based on the query's type and 
the mix of queries being executed by the database system. 
Q-Cop's model and \bouncer's response time estimates have similar formulations, and their decision
processes are alike too. 
But, rather than relying on a
model trained offline, \bouncer estimates percentile response times
based on recent processing times and the number of queries in the queue at the time. 
Besides, Q-Cop's model only captures average execution times; hence, it cannot enforce latency SLOs
defined in terms of percentiles.
Another prediction-based approach is ActiveSLA~\cite{activesla2011}, which makes acceptance
decisions to improve the profit of database-as-a-service providers while considering the service
level agreements (SLA) with their clients. 
It uses a non-linear classification model that considers the query types and mix, among other
features, to estimate the probability of a new query meeting its deadline.
Then, based on the probability, ActiveSLA decides whether to admit the query with a profit
optimization objective, with the expected profit derived from the SLA.
In contrast, \bouncer has no notion of profit and does not seek to maximize it. 
Unfortunately, ActiveSLA's overhead is much (1000x) higher than \bouncer's and becomes prohibitive
for systems like \liquid offering millisecond-scale response times.
In addition, ActiveSLA continually rebuilds the model 
to reduce the human effort on model retraining. 
But, for production data systems like \liquid, queries are regularly added and retired, altering the
query mix. 
Hence, ActiveSLA's model, as well as Q-Cop's, would need retraining more often than their authors
anticipate, not only when the database or hardware configurations change.  
For \bouncer, model retraining is not a concern.

\bouncer's overhead is low for millisecond-scale queries, which is key for request-by-request
acceptance decisions. 
Similarly, but at the operating-system level, MittOS~\cite{mittos2017} computes latency estimates
per IO operation in few $\mu$s or less, and embraces quick rejections for IOs that cannot be served
by a deadline.
Built within the storage stack,
it exposes a SLO-aware interface to help
reduce IO completion time for data-parallel applications. 
MittOS receives latency deadlines from applications and requires white-box knowledge 
of devices and resources, whereas \bouncer adopts a black-box approach and its latency SLOs are
percentiles.
MittOS neither considers call types nor tackles starvation, but applications could do so since 
IO calls are given individual deadlines.  
Targeting microsecond-scale remote procedure calls, Breakwater~\cite{breakwater2020} implements a
credit-based admission control scheme with demand speculation and overcommitment handling.
While offering fast rejections, it is oblivious to query types and only considers queue wait
times to determine the credits to be distributed distribution among clients.
Instead, \bouncer considers queue wait times and processing times to estimate percentile response
times for queries.

\section{Conclusion and Future Work}
\label{sec:conc}

We presented \bouncer, an admission control policy that makes query-by-query decisions based on
latency SLOs defined in terms of percentile response times.
It combines inexpensive approximations of queue wait time and processing time to estimate
percentile response times for each query, and compares the estimates with the objective
values to decide to accept or reject the query. 
It takes into account that production workloads often include various types of queries with
different latency characteristics, and allows assigning SLOs per query type.
In addition, we studied two starvation avoidance strategies (\aallow and \htu) that
supplement \bouncer's basic formulation to prevent query types from
systematically receiving no service.
We also conducted an extensive evaluation of \bouncer and its starvation-avoiding variants against
several in-house policies in simulation and on \liquid graph database. 
A summary of our experimental results is given in \S\ref{sec:result_summary}. 

As part of our \emph{future work}, 
we plan to evaluate alternative formulations for \bouncer that 
use higher-order percentile response times, 
apply different logical expressions for decision making,
and 
update processing time histograms in a sliding window, instead of non-overlapping windows.
We also intend to extend \bouncer to support queries served based on priorities
(rather than in FIFO order), and will consider adapting it to other query
scheduling disciplines. 

\bouncer is currently susceptible to cold starts and lulls in traffic because its decisions making
relies on per-query-type processing time histograms, which are initially in a blank state and become
empty when queries of certain types stop coming for some time. 
We expect \bouncer to handle these issues by itself, with no need of supplementary infrastructure and
no extra burden on system's operators. 
Appendix~\ref{sec:cold_start} discusses our plan in this regard.

Finally, we are interested in evaluating \bouncer against other policies in the literature as
well as its applicability within more general overload management solutions,
such as Quorum~\cite{quorum2005} (in its selective dropping module) and
LinkedIn's Hodor~\cite{hodor2022}.

\balance
\bibliographystyle{ACM-Reference-Format}
\bibliography{\papername}
\begin{acks}
We thank LinkedIn's graph team for their support and valuable feedback.
Special thanks to 
Andrew Carter for the \af policy (\$\ref{sec:acceptrate}), 
SungJu (Joe) Cho for the \mqwt policy (\S\ref{sec:maxqueuewaittime}), 
and 
Roman Averbukh for improving the \af policy and \liquid's admission control framework.
\end{acks}
\begin{appendices}
\section{Handling cold starts}
\label{sec:cold_start}

Like other measurement-based techniques, \bouncer suffers from the ``cold start'' syndrome because
its decision making relies on per-query-type processing time histograms, which are initially in a
blank state.
Note that different query types experience the problem independently, and the time they need to warm
up depends on 
when queries begin to arrive, 
at what pace, and 
whether they come continuously or sporadically -- aspects out of the policy's control. 
For instance, queries of popular types may eagerly arrive in continuous streams immediately after
the system starts, prompting the histograms to get filled early; by contrast, other query types
may arrive sporadically long after the system started, causing the histograms to become populated
at a later time. 
In this section, we discuss alternative solutions and our plan to handle cold starts in
\bouncer.

A possible solution to this problem is to warm up the system, before serving live traffic, with
sampled production queries amply representative of the different query types (like we did in
\S\ref{sec:impeller_results}).
But, it has practical implications. 
Every system's installation (\eg dozens of \liquid clusters per data center) must be warmed up right
after being deployed, which occurs often under continuous integration and deployment (CI/CD).  
This solution thus requires an ancillary software infrastructure, arguably part of the CI/CD
pipeline, that samples production traffic, coordinates warm-up executions, and issues sampled
queries to newly deployed system's installations. 
It also imposes additional burden on the system's operators since they need to manage this
extra piece of infrastructure and ensure that the query types in production are well
represented by the sampled queries.
Alternatively, one could think about deploying the system along with pre-populated histograms
containing query processing times from previous installations. 
But, besides needing mechanisms to capture, store, and redeploy histograms, this solution assumes
that past histograms remain representative of the queries' performance across versions of the
system, which is generally not true. 
The above solutions, which require supplementary infrastructure to handle cold starts, hinder
\bouncer's usability due to the added complexity.  
For that reason, we plan to adopt a solution in which \bouncer deals with cold starts by itself.

In our preferred solution, each query type goes through a \emph{warm-up phase}.
During this transitory phase, \bouncer lets queries in, possibly with some leniency, until the
histogram gets sufficiently populated, and once that happens the policy switches to normal operation
for the warmed-up query type.
The question is then: \emph{How should \bouncer operate during the warm-up phase?}
\bouncer includes a \emph{general histogram} where it stores the processing times of queries
regardless of their types.  
Thus, when the policy receives a query and finds out that the corresponding histogram is not
sufficiently populated, it gets 
the mean, 50th- and 90th-percentile processing times ($pt_{mean}$, $pt_{p50}$, and $pt_{p90}$) 
from the general histogram, and decides to admit or reject the query based on these values and the
latency SLOs for the \texttt{default} (catch-all) query type.
We also choose this solution because it is simple to implement and well-aligned with the \aallow
strategy (\S\ref{sec:acceptance_allowance}), which helps fill quickly the general and per-query-type
histograms. 

In addition, there is a related issue with queries following intermittent patterns. 
When queries of a given type stop coming for some time, the corresponding histogram may be 
1)~\emph{replaced by an empty one}, turning ineffective as it goes back to the initial blank state, 
or 
2)~\emph{retained}, despite becoming stale and its effectiveness possibly decaying with time. 
In this case we prefer stale data to no data; thus, we choose to retain the histograms when the
query counts are below a threshold.

Our solution is under development and its evaluation is left as future work.

\section{Other Practical Considerations}
\label{sec:considerations}

Here we discuss some aspects related to the use of \bouncer.

\subsection{Choosing percentiles for latency SLOs}
\label{sec:choosing_percentiles}

As discussed in \S\ref{sec:slo_policy}, \bouncer can be easily modified to support one, two,
or more percentile response times as objectives.
Then, the natural question is: \emph{What percentiles should we use in our latency SLOs?}
That decision depends on multiple factors, such as 
the characteristics of the system being considered, its workload (mix of queries and traffic volumes), and 
the percentile latencies the system's operators monitor and report.
But, one guiding principle is the \emph{stability} of the selected percentile values over time.

In this paper, we use the 50th- and 90th-percentile response times to indicate our latency
objectives ($SLO_{p50}$ and $SLO_{p90}$).
We chose them because, besides being commonly used to specify latency requirements, our
experience with \liquid shows that the 50th- and 90th-percentile processing times ($pt_{p50}$ and
$pt_{p90}$) observed by the brokers are more stable than $pt_{p99}$. 
Shards (as well as brokers) run a front-end component written in Java, and 
garbage collection pauses regularly cause relatively high $pt_{p99}$.
When a query type's histogram stores an elevated $pt_{p99}$ (\ie close to or larger than
$SLO_{p99}$), most of the queries of this type will be rejected in the next time interval until the
histogram is updated.
Instead, we found $pt_{p50}$ and $pt_{p90}$ to be less susceptible to garbage collection stalling.
 
\subsection{\bouncer's configuration effort}
\label{sec:conf_effort}

This is an aspect of practical importance because system operators prefer easy-to-configure policies
to simplify their duties. 
Since latency objectives under \bouncer are defined in terms of percentile response times (\eg
$SLO_{p50}$ and $SLO_{p90}$), its configuration comprises determining such percentiles for the query
types.  
But, this task is often necessary regardless of the admission control policy in use. 
The reason is that prospective and current customers of an online data system typically want to know
the response time for their queries and whether the system is able to meet their latency
requirements, especially at peak traffic load. 
Getting this information usually involves empirically evaluating the queries' performance under
realistic conditions (\eg with actual production traffic).
Thus, being part of the duties the system's operators need to carry out, obtaining the queries'
response time percentiles often requires no extra effort in reality.

Effort is also devoted to setting and maintaining the response time SLOs.
At the extreme, \bouncer allows having an SLO setting (\eg $SLO_{p50}$ and $SLO_{p90}$) per query type. 
Imagine being responsible for making sure that the individual SLOs for 100 query types in production are
properly set. 
That seems excessive when compared to using a utilization-centric policy, like \af, with a single
key configuration parameter: the maximum utilization limit.
In practice, however, the configuration effort is generally acceptable because multiple query types
often share the same SLO. 
We have seen ratios as high as 20:1, and our evaluation (\S\ref{sec:eval}) exemplifies a scenario
with an 11:1 ratio.
Such ratios suggest that operators can establish a manageable sized set of SLOs and assign
each SLO to multiple query types, effectively grouping queries into classes
(reminiscent of quality of service classes).  
 
In addition, \bouncer includes a \texttt{default} SLO setting for queries with unrecognizable type.
By setting permissive latency values in the default SLO, the system can serve new queries with no
declared type, and the operator can add the new query types and SLOs to the policy's configuration
later. 
Hence, the default SLO can help not only reduce the configuration effort involved in testing new
queries, but also avoid hindering their onboarding.

\end{appendices}

\end{document}